\documentclass[reprint,amsmath,amssymb,nofootinbib,aps,pra]{revtex4-2}

\usepackage{graphicx}
\usepackage{dcolumn}
\usepackage{bm}
\usepackage[breaklinks=true,colorlinks=true,linkcolor=blue,urlcolor=blue,citecolor=blue]{hyperref}
\usepackage{multirow}
\usepackage[dvipsnames]{xcolor}
\usepackage[pscoord]{eso-pic}
\usepackage[normalem]{ulem}
\usepackage{slashed}
\usepackage{makecell}

\newcommand{\placetextbox}[3]{
	\setbox0=\hbox{#3}
	\AddToShipoutPictureFG*{
		\put(\LenToUnit{#1\paperwidth},\LenToUnit{#2\paperheight}){\vtop{{\null}\makebox[0pt][c]{#3}}}
	}
}

\newcommand{\tquote}[1]{``#1''}

\def\mnras{MNRAS}

\begin{document}

\placetextbox{0.90}{0.97}{\small ULB-TH/21-08}

\title{Multi-interacting dark energy and its cosmological implications}

\author{Matteo Lucca}
\email{mlucca@ulb.ac.be}
\affiliation{Service de Physique Th\'{e}orique, Universit\'{e} Libre de Bruxelles, C.P. 225, B-1050 Brussels, Belgium}

\begin{abstract}
	In the effort to define the main features a successful solution to the Hubble tension should have, growing evidence has emerged pointing to the need for models able to modify the expansion (and possibly thermal) history of the universe both prior and after recombination, and that could thereby restore the overall concordance between early- and late-time observations without introducing nor worsening other cosmological tensions in the process. In precisely this spirit, here we consider a multi-interacting dark energy model with two complementary interaction channels: one with the dark matter and one with the photons. The former most significantly affects the matter dominated epoch as well as the very late universe, and has been shown to be able to significantly resolve the $S_8$ tension. The latter has been introduced in this work to allow for a larger and natural time dependence of the evolution of the universe since it mostly impacts the radiation dominated epoch as well as the temperature scaling of the photons, as extensively explained in the text. As a result, considering data from \textit{Planck}+BAO+Pantheon+KV450+DES (which can be combined since the $S_8$ value is lowered even neglecting the weak lensing data) we find that the significance of the $H_0$ tension only reduces to about 3.5$\sigma$, while that of the $S_8$ tension falls below the 2$\sigma$ level. Also, the statistical analysis we perform strongly favors $\Lambda$CDM in all considered cases. Overall, we conclude that the specific multi-interacting dark energy model considered here, despite its broad generality and very rich cosmological phenomenology, cannot successfully restore the aforementioned overall concordance between early- and late-time observations.
\end{abstract}

\maketitle

\section{Introduction}
Great theoretical and experimental efforts have been dedicated in the last decade to the understanding of the two most statistically significant and persistent tensions in the cosmological landscape, the Hubble tension and the $\sigma_8$ tension (see e.g., \cite{Perivolaropoulos:2021jda} for a recent review).

The former arises when comparing the inference of the $H_0$ value from early-time observations of the universe assuming the standard $\Lambda$CDM model, including most notably Cosmic Microwave Background (CMB) anisotropies \cite{Aghanim2018PlanckVI}, Baryon Acoustic Oscillation (BAO) and Big Bang Nucleosynthesis (BBN) data (see e.g., \cite{Schoeneberg2019BAO}), to late-time measurements of this quantity employing for instance the distance latter method \cite{Riess2019Large}. In particular, the cosmological observations probing the early universe suggest values around $H_0\simeq68$ km/(s Mpc), while the late-time counterparts prefer higher values of this quantity around $H_0\simeq73$ km/(s Mpc) (see e.g., Fig. 1 of \cite{DiValentino:2021izs} for a schematic summary). The fact that this discrepancy persists with an high significance across multiple observational strategies and data sets combinations as well as over many reanalysis and systematics studies of these data sets (see e.g., Sec. 2 of~\cite{DiValentino:2021izs} for a complete overview) makes of it one of the most interesting mysteries of modern cosmology.

Similarly, albeit with a lower statistical significance, the $\sigma_8$ tension also appears between the early-time inference and the late-time measurements, lead by weak lensing surveys such as the Kilo-Degree Survey (KiDS)~\cite{Hildebrandt:2018yau} and the Dark Energy Survey (DES) \cite{Troxel:2017xyo, Abbott:2021bzy}, of the amplitude of matter fluctuations, referred to as~$\sigma_8$ when taken at the benchmark scale of 8~Mpc/$h$. Since this parameter usually shares a strong degeneracy with the matter energy density $\Omega_m$, it is commonly rephrased in terms of the parameter $S_8=\sigma_8\sqrt{\Omega_m/0.3}$, which does account for the degeneracy. In this case, early-time probes suggest values around $S_8\simeq0.83$ \cite{Aghanim2018PlanckVI}, while the results of direct observations can vary quite significantly but consistently deliver values below ${S_8\simeq0.80}$ (see e.g., Tab. III of \cite{Perivolaropoulos:2021jda}).

Clearly, it is very intriguing to consider the possibility that these tensions could be the first hints for a cosmological model beyond $\Lambda$CDM, and indeed this speculative approach has motivated a plethora of models aiming at alleviating these discrepancies (see e.g., \cite{DiValentino:2020zio, DiValentino:2021izs} and \cite{DiValentino:2020vvd} for recent reviews of models addressing the $H_0$ and $S_8$ tensions, respectively). Nevertheless, in particular in the context of the Hubble tension, the abundance and complementarity of the many data sets available impose significant constraints on these models. For instance, the interplay between BAO and SNIa data seems to prefer $\Lambda$CDM over possible late-time modifications of the expansion history \cite{Poulin2018Implications}. Furthermore, these late-time attempts to address the $H_0$ tension would anyway be intrinsically unable to affect the sound horizon at recombination time, which however needs to be decreased by about 10 Mpc in order to successfully solve the $H_0$ tension~\cite{Bernal2016Trouble,Knox2019Hubble}. On the other hand, as pointed out in \cite{Jedamzik:2020zmd}, pure early-time solutions strongly affecting the matter density parameter introduce tensions with BAO or weak lensing data. Also, \cite{Wong:2019kwg, Krishnan:2020obg} brought forward evidence for a late-time variation of the $H_0$ parameter, favoring a late-time solution over the early-time alternatives.

Overall, there seems to be growing consensus around the idea that for a solution to the Hubble tension to be successful it needs to modify the expansion history of the universe just prior to recombination, but might require additional late-time modifications to the $\Lambda$CDM model to compensate for the new physics introduced at earlier times so that it would not introduce new tensions among the many data sets (see e.g., \cite{Vagnozzi:2021tjv} additional compelling arguments in this direction). Possibly, this late-time behavior would then also be able to address the $S_8$ tension. 

Precisely this philosophy has been recently employed in~\cite{Allali:2021azp}, where the authors develop a phenomenological dark sector with decaying dark energy (DE) and ultra-light axions (both known to be able to shift the $H_0$~\cite{Niedermann:2020dwg} and $S_8$~\cite{Hlozek:2014lca} values in the right direction, respectively) to significantly reduce the statistical significance of both tensions. Similarly, \cite{Becker:2020hzj} constructed a multi-interacting dark matter (DM) model where DM-dark radiation interactions can improve the $H_0$ tension and DM-photon interactions the $S_8$ tension. Although a compelling and complete microphysical origin of these dark sectors remains an open question, as the authors of the references point out, this approach remains a very important exercise to highlight the cosmological features a successful model might have to have in order to restore the concordance between \textit{all} early- and late-time observations. 

Following this idea, in this work we aim at obtaining a conceptually similar result to \cite{Allali:2021azp,Becker:2020hzj}, but with very different phenomenological features. We build on the results recently discussed in \cite{Lucca:2021dxo}, where is has been shown that allowing for an energy transfer from the DE to the DM can greatly reduce the significance of the $S_8$ tension, and extend this model by increasing the number of species the DE can interact with. In particular, in order to be able to smoothly affect the early as well as the late universe, we consider the case of DE-photons interactions.

Clearly, the inclusion of the interaction with the photons in this generalized setup implies a variety of interesting and non-trivial consequences. First of all, being the photons the dominating fluid prior to recombination, this interaction introduces an important model dependence of the pre-recombination history. In this way, the combination of DM and photon interactions provides a large and natural freedom in the time-dependence of the model, which can impact both the radiation dominated (RD) and matter dominated (MD) epochs (not requiring any redshift-dependent fine tuning, present for instance in models such as Early DE~\cite{Poulin2018Early, Smith2019Oscillating, Hill2020Early}). This dynamics could then in principle allow to overcome both the aforementioned difficulties of late-time models, such as the BAO and SNIa constraints as well as the discrepancy in the value of the sound horizon, and those raised in \cite{Jedamzik:2020zmd, Lin:2021sfs} against pre-recombination solutions of the $H_0$ tension. 

Moreover, photon interactions inevitably affect the temperature evolution of the universe, thereby impacting the cosmological history at a deeper level than when only DM-DE interactions are considered. In this way, for instance, it becomes in principle possible to affect the recombination history and thereby the redshift of photon decoupling, which ultimately determines the sound horizon and could play an important role in the context of the Hubble tension.

Although some features of this model have already been studied at background \cite{Jetzer:2010xe, Jetzer:2011kw} and thermodynamics~\cite{Lima:1995kd, Lima:2000ay} level, to our knowledge, a derivation of the cosmological perturbation equations has never been proposed and hence a full analysis of the cosmological constraints has never been conducted before. For instance, because of the limited number of observables considered in previous works, imposing independent constraints on both coupling parameters at the same time has never been possible. However, by deriving the full set of cosmological equations in this work we can go beyond this limit. Furthermore, also the role of the DE equation of state (EOS) parameter $w_x$ has not been fully explored yet, always being kept fixed to the standard value of $-1$ in previous works. Here we treat this quantity as a free parameter for sake of generality throughout the full calculations.

The paper is organized as follows. In Sec. \ref{sec: math} we review, update and extend the mathematical setup required to investigate the cosmological implications of the aforementioned interacting DE (iDE) model. Many additional details and full calculations are provided in the dedicated appendices. In Sec. \ref{sec: obs} we discuss the relation between the model's parameters and some key cosmological observables such as the CMB temperature power spectrum and the matter power spectrum, with particular emphasis on the physical origin of the deviations from $\Lambda$CDM. In Sec. \ref{sec: meth} we describe the numerical setup and cosmological data sets used to evaluate the model. In Sec. \ref{sec: res} we discuss the results obtained for different parameter choices and data set combinations. Finally, in Sec. \ref{sec: conc} we summarize the main features of the model, and conclude with final discussions and remarks.

\section{The mathematical setup}\label{sec: math}

In this section we will outline the many cosmological implications of the iDE model considered in this work. In particular, we will discuss the background evolution in Sec. \ref{sec: math_bg}, the modifications to the thermal history in Sec. \ref{sec: math_th}, and finally the corresponding perturbation equations are presented in Sec. \ref{sec: math_pt}. Detailed derivations of the main equations are proposed in several dedicated appendices. 

The concrete numerical examples and graphical representations employed in the section to facilitate the physical understanding of a particular aspect of the model are computed with the numerical setup discussed in Sec. \ref{sec: meth}, unless stated otherwise. Moreover, when referring to the standard scenario, we implicitly mean the $\Lambda$CDM model with parameter values fixed to the \textit{Planck}(+BAO) best fits reported in the last column of Tab. 2 of \cite{Aghanim2018PlanckVI}. If not explicitly stated differently, these will also be the values we will assume for the standard six cosmological parameters in the aforementioned numerical examples.

\subsection{Background}\label{sec: math_bg}
At the background level the main difference with respect to the standard $\Lambda$CDM model is that, being the three fluids coupled, the individual energy densities are not conserved singularly any more but are instead related via an energy exchange term $Q$, reading \cite{Jetzer:2010xe, Jetzer:2011kw}
\begin{align}
	& \label{eq: rho_c} \dot{\rho}_{c}+3H\rho_{c}=(1-\epsilon)Q\,, \\
	& \label{eq: rho_g} \dot{\rho}_\gamma +4H\rho_\gamma =\epsilon Q\,, \\
	& \label{eq: rho_x} \dot{\rho}_x +3H\rho_x (1+w_x) =-Q\,.
\end{align}
Here $\epsilon$ rules the amount of energy transferred from the DE to the DM or the photons, or \textit{vice versa} (depending on the sign of $Q$), and is bound between 0 (only DE-DM interactions) and 1 (only DE-$\gamma$ interactions), while $w_x$ is the DE EOS parameter\footnote{Note that, in the spirit outlined in the introduction, these equations are purely phenomenological and parametrize in a broad way the possible interactions between the various fluids. Should such setup be favored by data, a careful derivation from a fundamental theory would become the necessary next step of our investigation. Nevertheless, we will not touch upon this aspect of the model within this work (see however \cite{Jetzer:2010xe, Jetzer:2011kw} and references therein for possible efforts in this direction).}. For sake of brevity, it is useful to introduce the effective DE EOS parameter $w_{{\rm eff},x}$ defined as
\begin{align}\label{eq: w_eff_x}
	w_{{\rm eff},x} = w_x + \frac{Q}{3H\rho_x}\,,
\end{align} 
which can be derived from Eq. \eqref{eq: rho_x}. Note that we have implicitly assumed $w_c=0$ and $w_\gamma=1/3$ for the DM and photon EOS parameters\footnote{As clear from Eqs. \eqref{eq: rho_c}-\eqref{eq: rho_g}, even if the standard EOS parameters $w_c$ and $w_\gamma$ are fixed to 0 and $1/3$, respectively, the DM and photon EOSs can still vary due to an effective additional contribution to the EOS parameter proportional to $Q/(3H\rho_i)$, where $i=c,\gamma$. As we will see later, these contributions are directly proportional to the parameter $\xi$, which is left free.}, respectively, while the DE EOS parameter is treated as a free parameter for sake of generality.

As discussed in e.g., \cite{Lucca:2021dxo}, assuming $\omega_x\simeq-1$ one can use Eq. \eqref{eq: rho_x} to define the energy transfer function as
\begin{align}\label{eq: C_x}
	Q=-\frac{\dot{\Lambda}}{8\pi G}\,,
\end{align}
recalling that $\rho_x=\Lambda/(8\pi G)$. This then takes the common form
\begin{align}\label{eq: C_x 2}
	Q=\xi H\rho_x
\end{align}
under the assumption that $\Lambda$ is not a constant as in the $\Lambda$CDM model, but follows instead the simple but arbitrary redshift dependence
\begin{align}\label{eq: lambda}
	\Lambda=\Lambda_0(1+z)^\xi\,,
\end{align}
where $\Lambda_0=\Lambda(z_0)=3H_0^2\, \Omega_\Lambda$ and $\xi$ is an additional free parameter of the model. With Eq. \eqref{eq: C_x 2} it follows from Eq. \eqref{eq: w_eff_x} that $w_{{\rm eff},x}=w_x+\xi/3$.

Once the functional form of $Q$ is known, it is possible to analytically solve (see App. \ref{sec: app bg}) the background evolution of the energy densities as
\begin{align}
	\nonumber \rho_c= & \, \rho_{c,0}(1+z)^3 \\
	& \label{eq: rho_c 2} +(1-\epsilon)\xi\frac{\rho_{x,0}(1+z)^3}{3w_{{\rm eff},x}} [1-(1+z)^{3w_{{\rm eff},x}}]\,, \\
	\nonumber\rho_\gamma= & \, \label{eq: rho_g 2} \rho_{\gamma,0}(1+z)^{4} \\
	& +\epsilon \xi\frac{\rho_{x,0}(1+z)^{4}}{3w_{{\rm eff},x}-1} [1-(1+z)^{3w_{{\rm eff},x}-1}]\,, \\
	\label{eq: rho_x 2}\rho_x= & \, \rho_{x,0}(1+z)^{3(1+w_{{\rm eff},x})}\,.
\end{align}
As expected, for $\epsilon=0$ we obtain Eqs.~\text{(2.4)-(2.5)} of~\cite{Lucca2020Shedding}, while for $w_x=-1$ Eqs. (2.9)-(2.11) of \cite{Jetzer:2011kw} (assuming $\gamma=4/3$). From the equations for the DM and the photons, we immediately obtain two conditions on $\xi$ and $\epsilon$ by imposing that the energy densities stay always positive. Indeed, neglecting the extra redshift dependences (which vanish at high redshifts), assuming $w_x\simeq-1$ and using $\Omega_{c}\simeq0.26$, ${\Omega_{\gamma}\simeq 5.4 \times 10^{-5}}$ and $\Omega_{x}\simeq0.69$, from Eq. \eqref{eq: rho_c} we get that $(1-\epsilon)\xi<1$, while from Eq. \eqref{eq: rho_g} we have that $\epsilon \xi<3\times 10^{-4}$. This means that, as long as $\xi$ is negative, every value of $\epsilon$ is in principle allowed, while when $\xi$ is positive it has to be lower than approximately 1 with $\epsilon=0$ and than $3\times10^{-4}/\epsilon$ with $\epsilon\neq 0$.

\begin{figure*}
	\centering
	\includegraphics[width=\columnwidth]{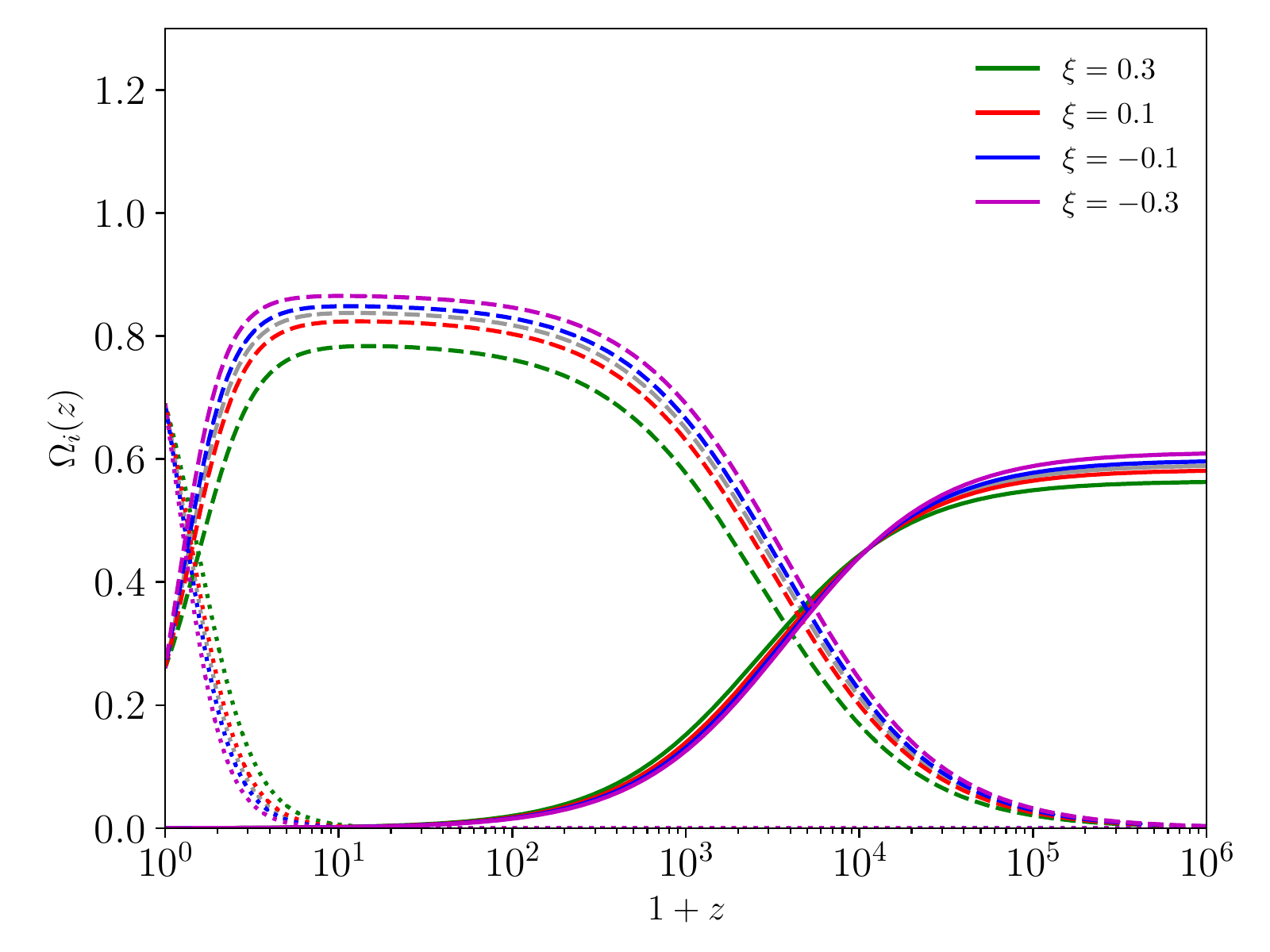} 
	\includegraphics[width=\columnwidth]{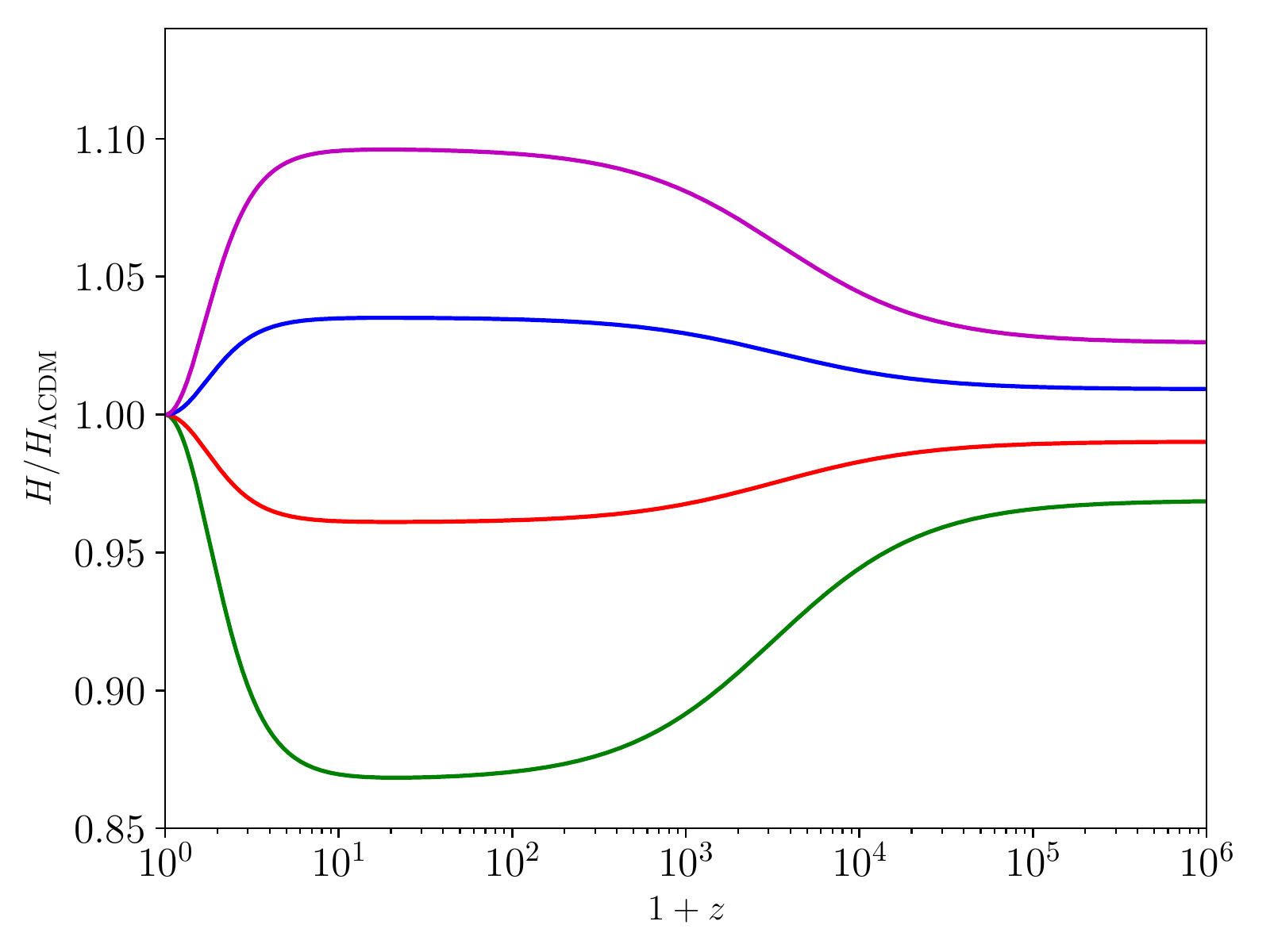}
	\\
	\includegraphics[width=\columnwidth]{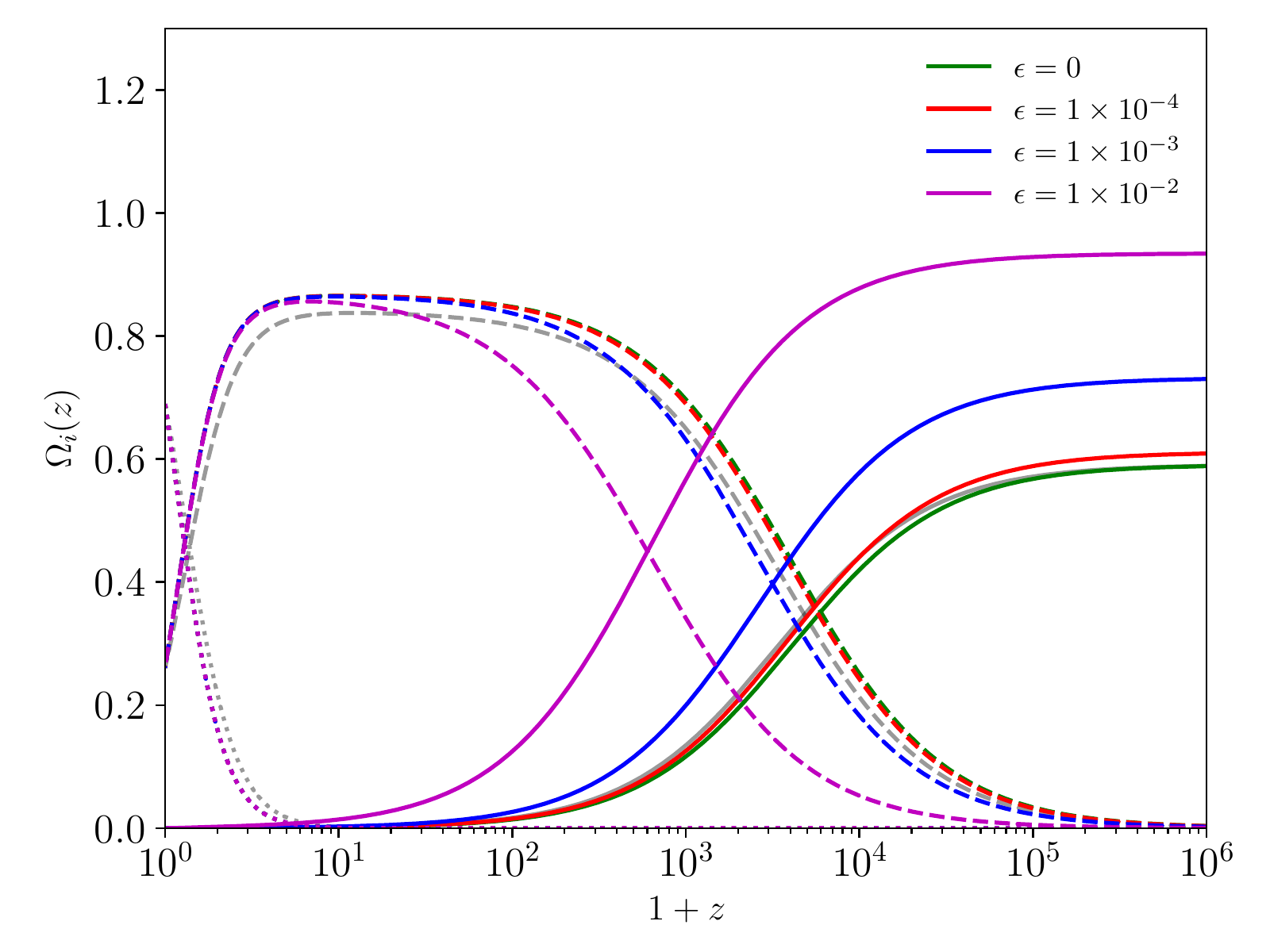}
	\includegraphics[width=\columnwidth]{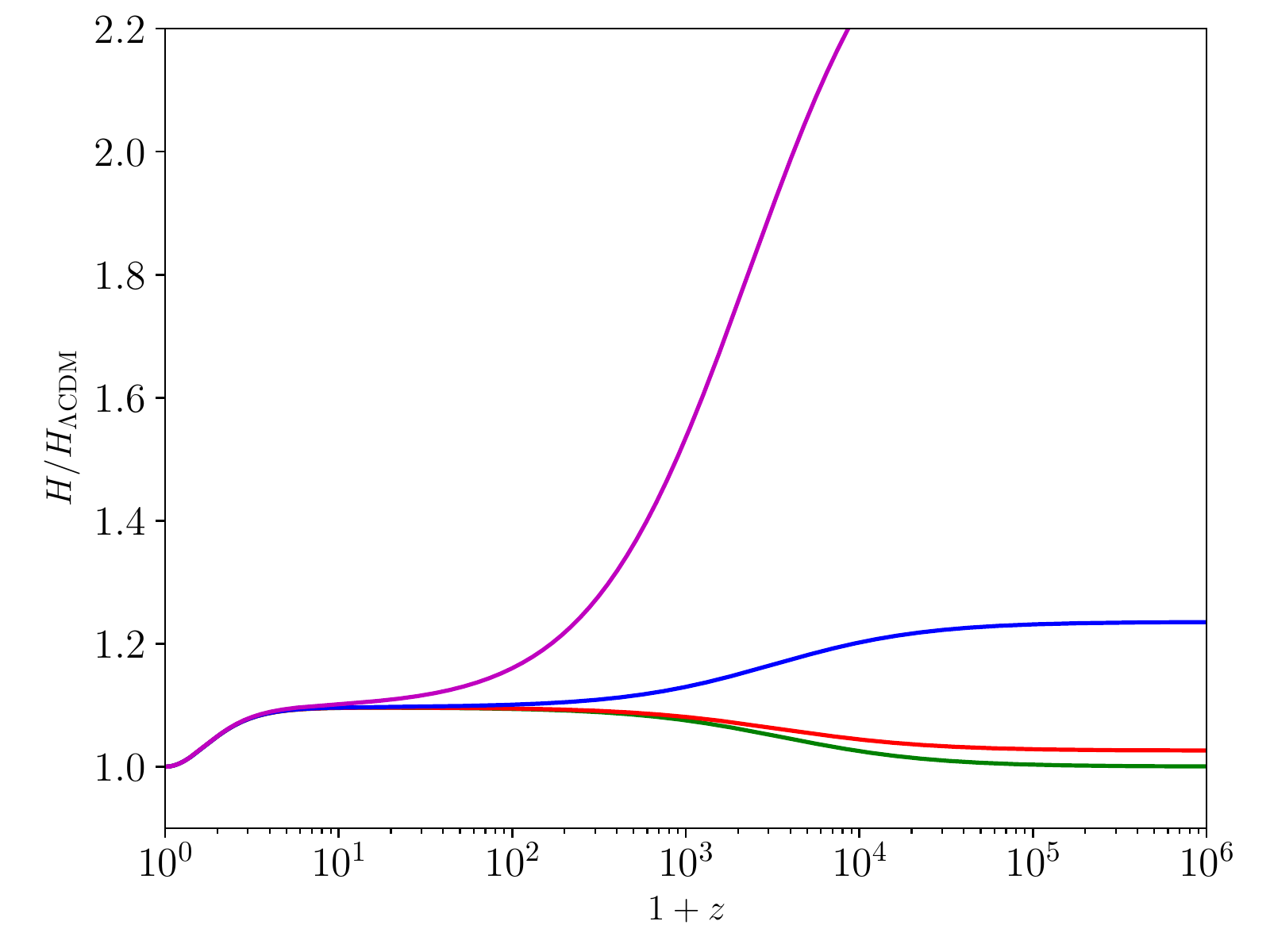}
	\\
	\includegraphics[width=\columnwidth]{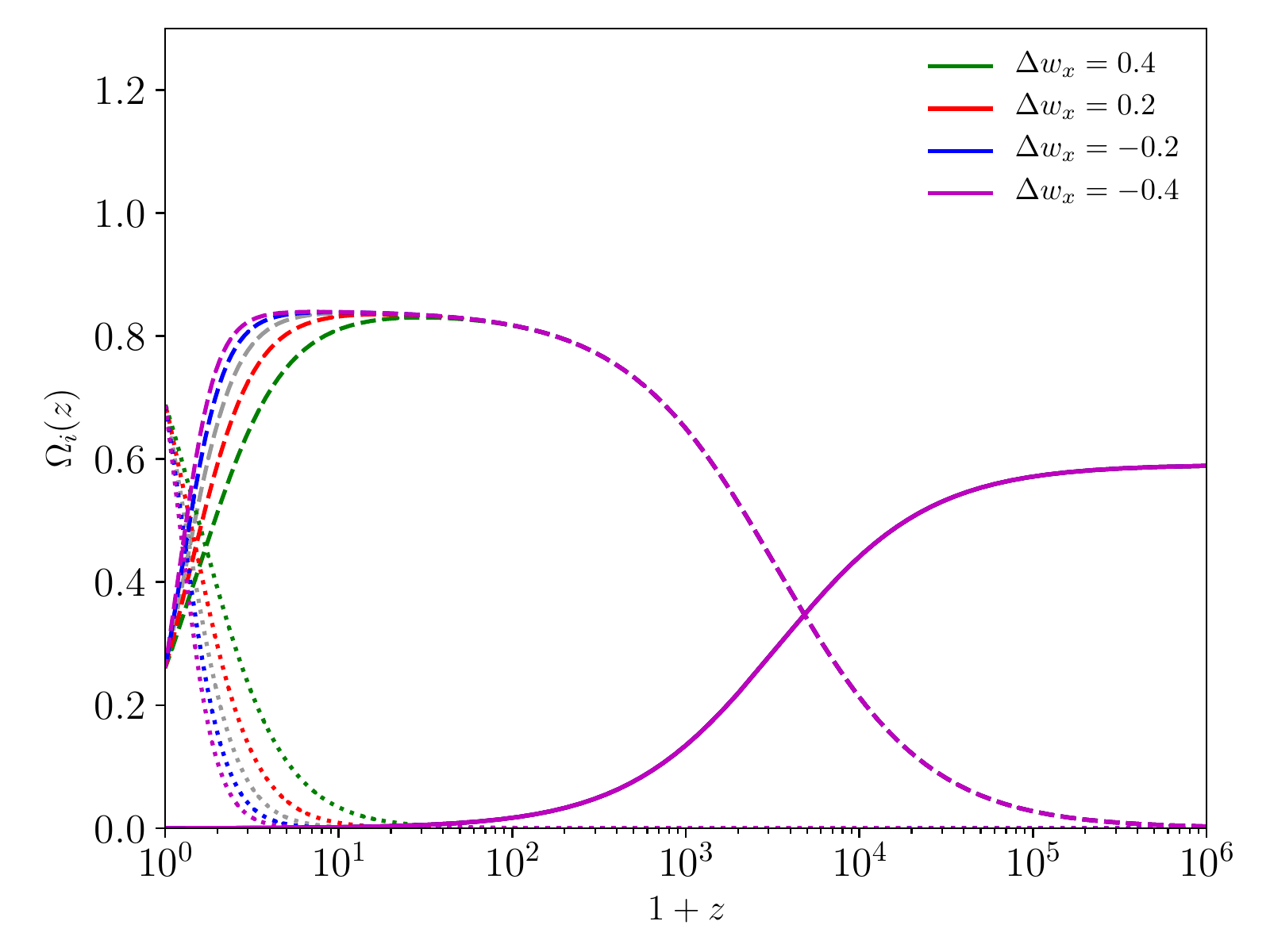}
	\includegraphics[width=\columnwidth]{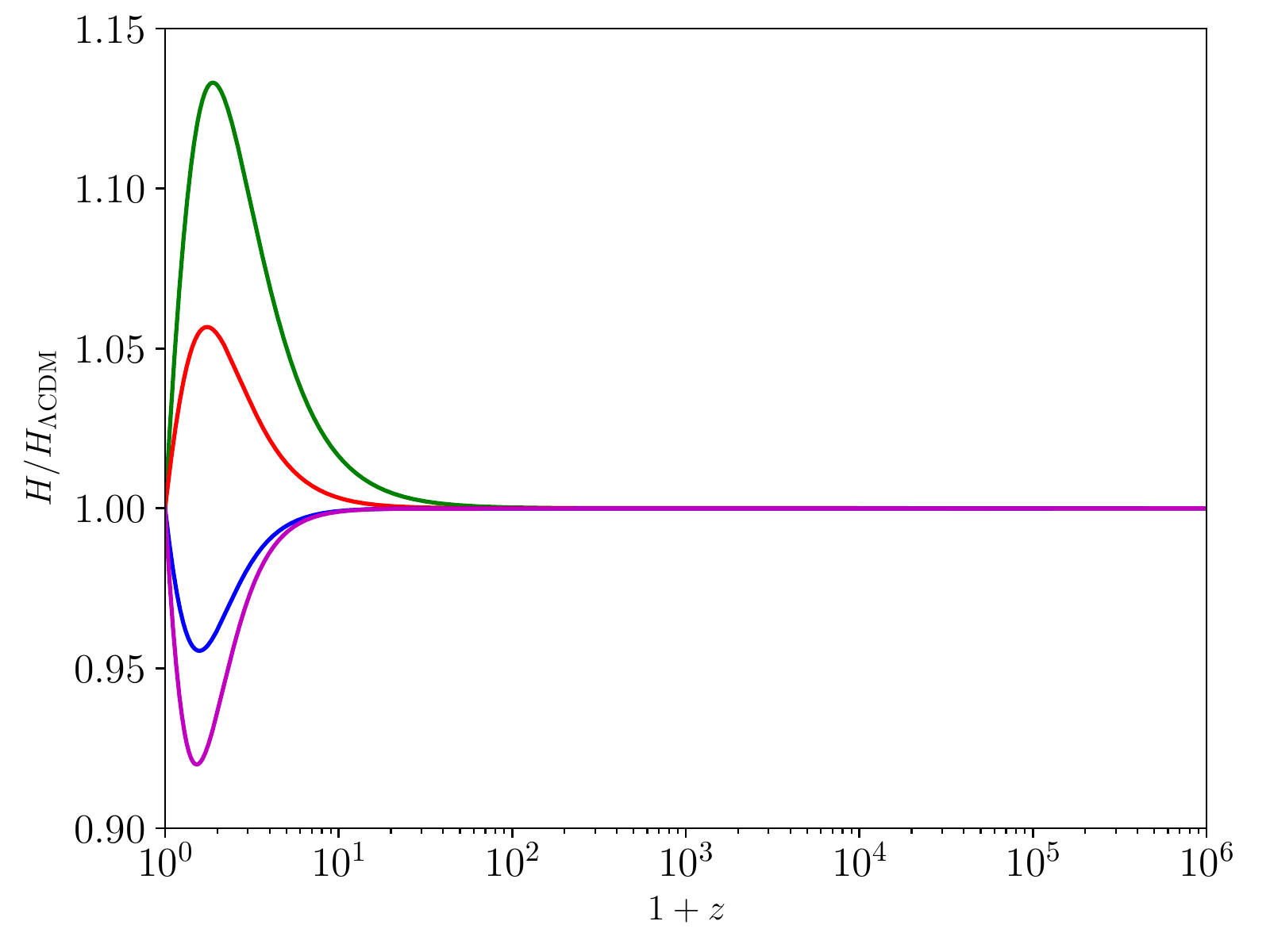}
	\caption{\textit{Left panels}: Effect of varying the free parameters of the iDE model considered within this work ($\xi$, $\epsilon$ and $w_x$) on the evolution of the DE (dotted lines), DM (dashed lines) and photon (solid lines) energy densities (expresses as $\Omega_i=\rho_i/\rho_{\rm crit}$, where $\rho_{\rm crit}(z)$ is the critical energy density and $i=x,c,\gamma$). The gray lines represent the $\Lambda$CDM case. Unless stated otherwise, we assume $\xi=0$, $\epsilon=1\times10^{-4}$ and $w_x=w_{x,\Lambda\text{CDM}}=-1$. From top to bottom we focus specifically on the role of $\xi$, $\epsilon$ with $\xi=-0.3$ and $w_x$. For sake of clarity, the latter is referred to as $\Delta w_x=w_x-w_{x,\Lambda\text{CDM}}$. \textit{Right panels}: Same as in the left panels, but considering the effect of the free parameters on the Hubble parameter $H(z)$ (normalized to the $\Lambda$CDM value $H_{\Lambda{\rm CDM}}$).}
	\label{fig: evolution_O}
\end{figure*}

A graphical representation of these functions is shown in the left panels of Fig. \ref{fig: evolution_O}, where we investigate the impact that the different free parameters have on them. For sake of completeness and clarity, in the right panels of the same figure we show the corresponding modifications to the Hubble rate $H$ (normalized to the $\Lambda$CDM evolution). In particular, focusing first of all on the top panel, it is interesting to notice that the role of $\xi$ is that of scaling up or down the DM and photon energy densities (depending on whether $\xi$ is negative or positive), with an opposite effect on the DE. Furthermore, since we have implicitly assumed a small value of $\epsilon$ for all cases  ($\epsilon=1\times10^{-4}$), the change in the DM energy density is stronger than the one in the photon energy density (the former being proportional to $(1-\epsilon)$ and the latter to $\epsilon$).

As clear from the middle panel of the figure, however, this behavior changes when $\epsilon$ increases. Indeed, for $\xi$ fixed to the fiducial value of $-0.3$, already with $\epsilon=0.001$ we observe that $\Omega_\gamma$ is significantly more affected by the non-standard interactions than $\Omega_c$ for approximately $z\geq10^3$. For higher values of $\epsilon$ the Hubble rate quickly diverges towards higher values during RD. To this point, it is also worth noting that, as mentioned before, in the case where $\xi$ is positive (with the same absolute value of 0.3), the value of $\epsilon$ cannot be increased arbitrarily without ending up with a negative photon energy density. Therefore, in this case the Hubble rate does not diverge to extreme values, but rather it decreases until it reaches zero already for values of $\epsilon$ of the order of $10^{-3}$. Furthermore, importantly, the effects that these two parameters, $\xi$ and $\epsilon$, have on the evolution of the energy densities are not degenerate. Finally, in the bottom panel of the figure we present some representative examples of the effect of varying $\Delta w_x=w_x-w_{x,\Lambda\text{CDM}}$. In this case, the resulting modifications with respect to $\Lambda$CDM are only at very late times and there is no clear degeneracy between $w_x$ and the other free parameters.

\begin{figure}
	\centering
	\includegraphics[width=\columnwidth]{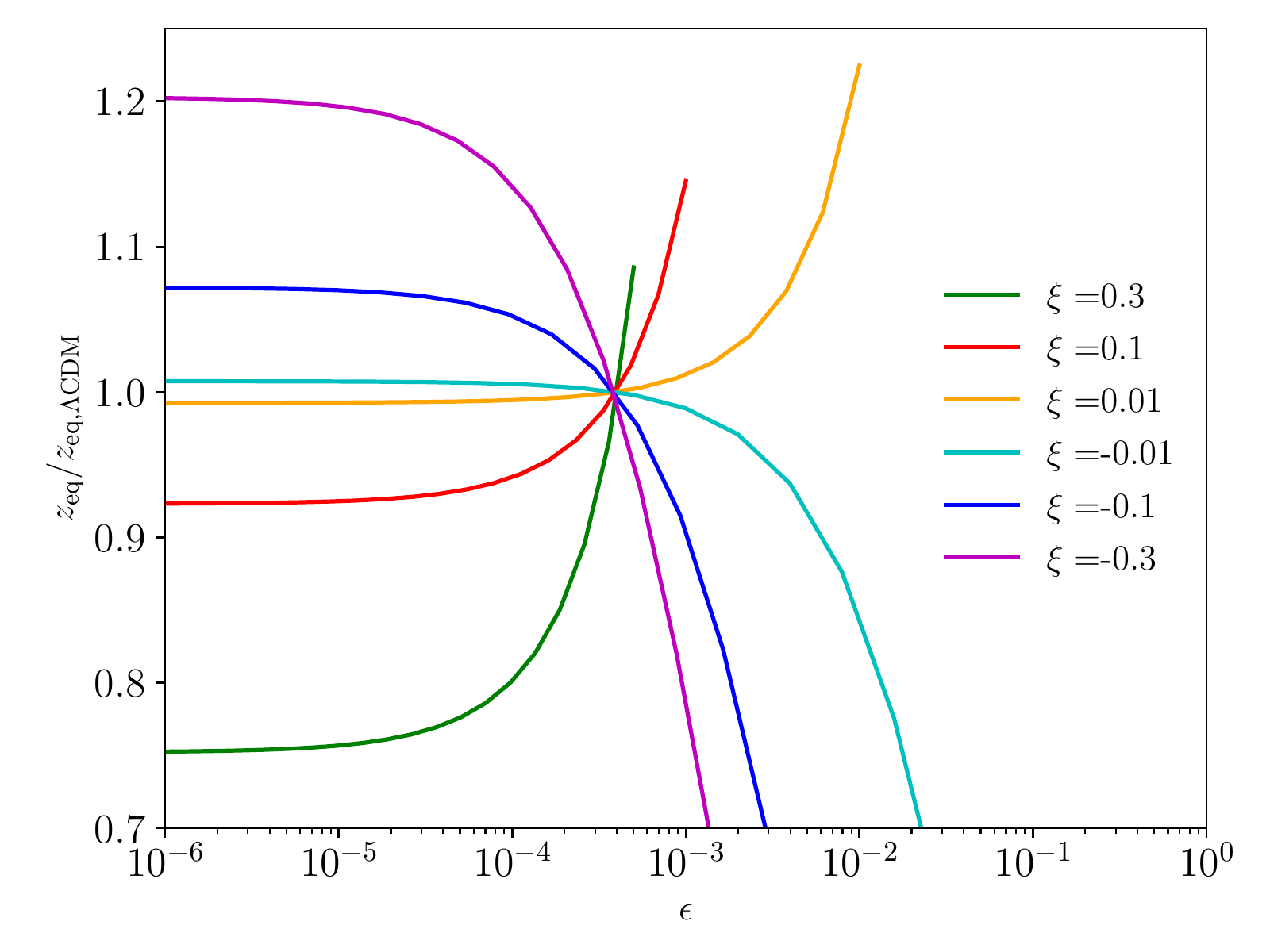}
	\caption{Dependence of the redshift of matter-radiation equality $z_{\rm eq}$ (normalized to the $\Lambda$CDM value $z_{{\rm eq},\Lambda{\rm CDM}}=3387$) on the free parameters of the model $\xi$ and $\epsilon$. The lines are interrupted whenever an assumption required by CLASS is violated.}
	\label{fig: z_eq}
\end{figure}

One important consequence of this modified background evolution is that also the redshift of matter-radiation equality $z_{\rm eq}$ changes. We show the extent of this variations in Fig. \ref{fig: z_eq} for different values of $\xi$ and as a function of $\epsilon$. As expected from the previous discussion, for very low values of $\epsilon$ (below $1\times10^{-4}$) $z_{\rm eq}$ is almost only depending on $\xi$. However, as soon as $\epsilon$ increases enough the amount of photon energy density rapidly grows (see the middle right panel of Fig. \ref{fig: evolution_O}), so that $z_{\rm eq}$ quickly increases or decreases depending on whether $\xi$ is positive or negative.
\begin{figure}
	\centering
	\includegraphics[width=\columnwidth]{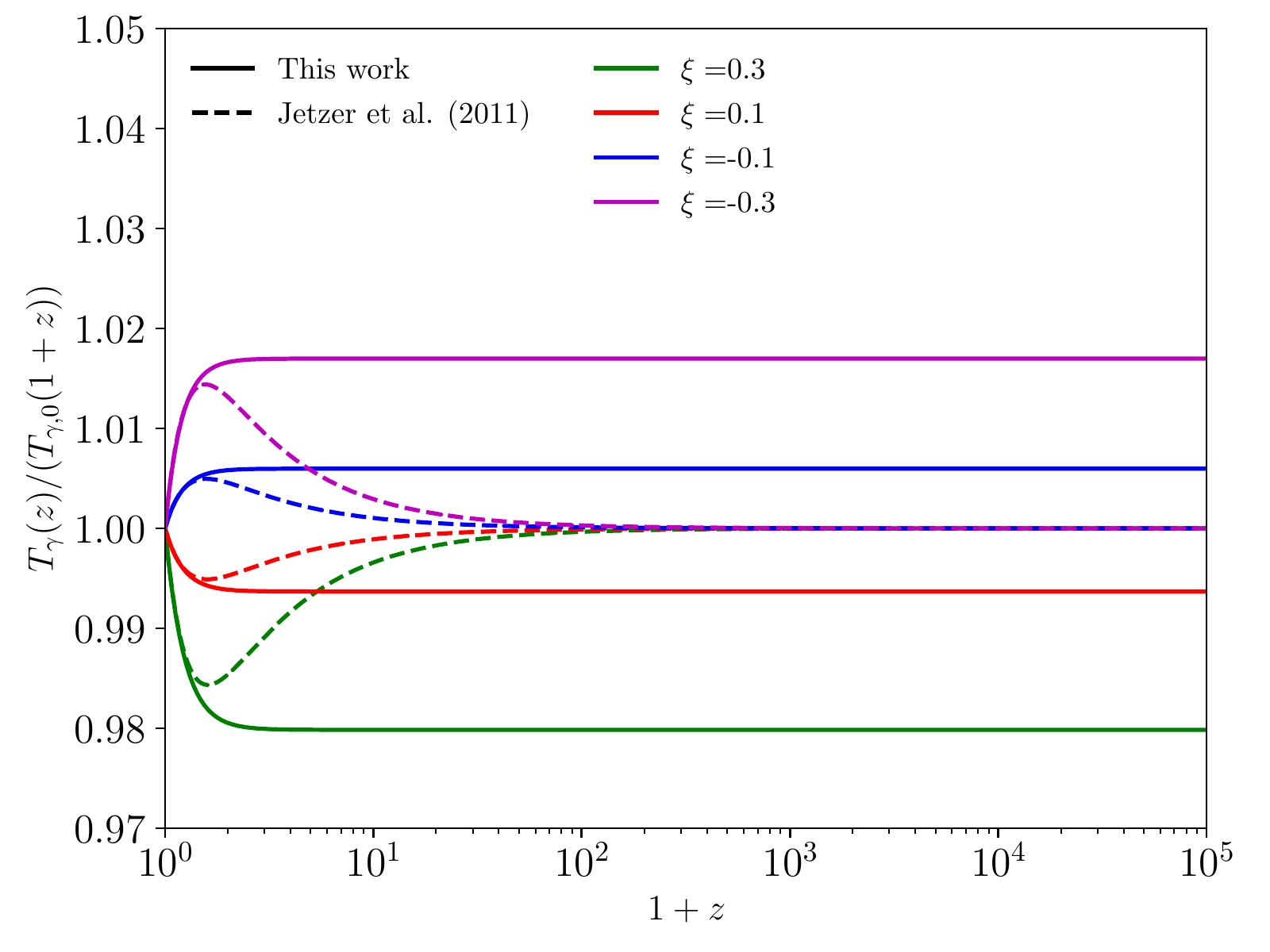}
	\caption{Evolution of the CMB photon temperature (normalized to the $\Lambda$CDM scaling) as function of redshift. The solid lines represent the result found in this work, while the dashed lines correspond to the one derived in \cite{Jetzer:2011kw} (without any assumption on $\tilde{\epsilon}$). The different colors correspond to different values of $\xi$. Note that in Eq.~\eqref{eq: T_g} there is an almost perfect degeneracy between $\xi$, $\epsilon$ and $w_x$ so that we do not vary the latter two.}
	\label{fig: evolution_T}
\end{figure}

\subsection{Thermodynamics}\label{sec: math_th}
In order to understand the thermodynamical implications that assuming the aforementioned DE interactions has on the cosmological model, it is useful to reformulate Eqs. (\ref{eq: rho_c})-(\ref{eq: rho_x}) in terms of the particle number densities $n_i$ and source terms $\psi_i$, i.e.,
\begin{align}\label{eq: dot_n}
	\dot{n}_i+3n_i H =\psi_i\,,
\end{align}
where $i=$ DM or $\gamma$. In this way, it then becomes possible to analytically solve the generalized temperature evolution equation for the photons \cite{Lima:1995kd, Lima:2000ay}
\begin{align}\label{eq: dot_t}
	\nonumber \frac{\dot{T_\gamma}}{T_\gamma}= & \left(\frac{\partial p_\gamma}{\partial \rho_\gamma}\right)_n\frac{\dot{n}_\gamma}{n_\gamma} \\ & -\frac{\psi_\gamma}{n_\gamma T_\gamma(\partial \rho_\gamma/\partial T_\gamma)_n}\left(p_\gamma+\rho_\gamma-\frac{n_\gamma \epsilon Q}{\psi_\gamma}\right)
\end{align}
to find (see App. \ref{sec: app ther})
\begin{align}\label{eq: T_g}
	\nonumber T_\gamma = & T_{\gamma,0} (1+z)\left[1+\frac{\epsilon \xi}{3w_{{\rm eff},x}-1}\frac{\rho_{x,0}}{\rho_{\gamma,0}}\right. \\ & \hspace{2. cm} \times \left. (1-(1+z)^{3w_{{\rm eff},x}-1}) \right]^{1/4}\,,
\end{align}
only by enforcing the condition that the spectrum of the CMB photons remains that of a black body over the whole thermal history of the universe. Of course, this approximation is rather crude, in particular at late times when the thermalization of the CMB photon spectrum is very inefficient, and noticeable CMB spectral distortions~\cite{Chluba2011Evolution, Chluba2019Voyage, Kogut2019CMB, Lucca2019Synergy} are expected to arise \cite{Chluba:2014wda}. Nevertheless, we will follow this assumption here for sake of simplicity, and leave a more careful and complete treatment of the problem including CMB spectral distortions as a complementary probe for future work.

Note that Eq. \eqref{eq: T_g} differs from the solution found in Sec. II D of \cite{Jetzer:2011kw} (with $\gamma=4/3$). One reason for this is that, since in \cite{Jetzer:2011kw} only data points with redshifts below 5 were considered, the role of radiation could be neglected there. Also, in order to reduce the number of degrees of freedom (DOF) and simplify the analysis, in \cite{Jetzer:2011kw} it was assumed that the whole matter content of the universe was in form of DM. Although these approximations are quite accurate at low redshifts, here we go beyond these assumptions and Eq.~\eqref{eq: T_g} generalizes the results presented in \cite{Jetzer:2011kw} to arbitrarily high redshifts and to the case where $w_x$ is free to vary. 

A graphical comparison between the solution found in this work (solid lines) and the one derived in \cite{Jetzer:2011kw} (dashed lines) is proposed in Fig. \ref{fig: evolution_T}. There, we show the evolution of the photon temperature (normalized to the $\Lambda$CDM case) for different values of $\xi$ (which in the case of the temperature is almost perfectly degenerate with $\epsilon$). As clear from the figure, and as expected, the two cases correspond well at very low redshifts, but while the solution of \cite{Jetzer:2011kw} rapidly falls back to the $\Lambda$CDM scaling, the extended version of Eq.~\eqref{eq: T_g} preserves a constant offset. The physical interpretation of this effect is easy to understand. Indeed, the non-standard source term in Eq.~\eqref{eq: dot_T}, $\psi_\gamma/n_\gamma$, is proportional to $H\rho_x/\rho_\gamma$ or more simply $H/\rho_\gamma$ if $\xi$ is small and $w_x$ is close to its standard value. Then, if radiation is neglected, as in~\cite{Jetzer:2011kw}, $H\propto a^{-3}$ at early times and $\psi_\gamma/n_\gamma\propto a$, which quickly becomes negligible, so that the extra source term in the temperature evolution disappears and the standard scaling is recovered. However, by including the contribution from radiation in the Hubble parameter one has that $H\propto a^{-4}$ at early times and $\psi_\gamma/n_\gamma\propto a^0$, so that we obtain the aforementioned constant offset at high redshifts displayed in Fig.~\ref{fig: evolution_T}.

Another important point to bear in mind when comparing our results to the ones found in \cite{Jetzer:2011kw} is that in the reference the parameter $\epsilon$ has been redefined to ${\tilde{\epsilon}=\epsilon\rho_m/(\gamma \rho_\gamma)}$, where $\gamma=1+w_\gamma$. Furthermore, because of the limited amount of observables employed in~\cite{Jetzer:2011kw}, i.e., only the late-time evolution of the CMB temperature, and the almost perfect degeneracy shared between $\xi$ and $\epsilon$ in Eq. \eqref{eq: T_g}, in the reference the condition $\tilde{\epsilon}=1$ has been artificially assumed. One significant consequence of this assumption is that it introduces a redshift dependence in $\epsilon$ ($\epsilon \to \epsilon\propto (1+z)$) because of the different scalings of $\rho_m$ and $\rho_\gamma$. Although this effect does compensate for the incorrect scaling of the source term in the photon temperature at early times (which then becomes $\psi_\gamma/n_\gamma\propto \epsilon H/\rho_\gamma \propto a^0$), it changes the dynamics of the model in an inconsistent way with respect to the background equations. For this reason and for sake of generality, here we will not make any assumption on $\epsilon$ (or $\tilde{\epsilon}$), treating it as a purely free parameter. With the inclusion of the many different observational probes considered within this work (see Sec. \ref{sec: meth}) the problematic degeneracy between $\xi$ and $\epsilon$ mentioned in~\cite{Jetzer:2011kw} is not a problem here (see Sec. \ref{sec: res}).
\begin{figure}
	\centering
	\includegraphics[width=\columnwidth]{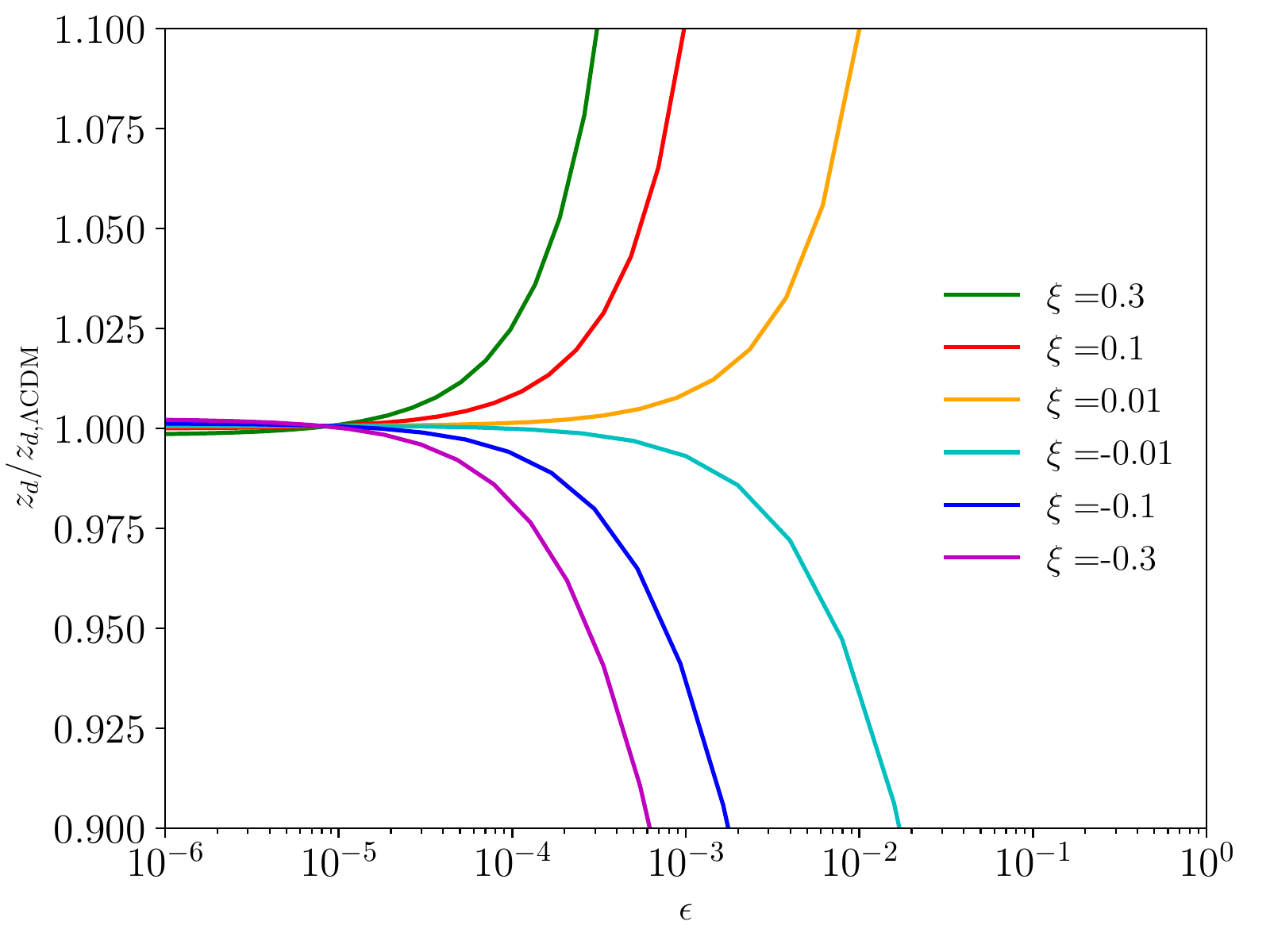}
	\caption{Dependence of the redshift of recombination $z_d$ (normalized to the $\Lambda$CDM value $z_{d,\Lambda{\rm CDM}}=1088$) on the free parameters of the model $\xi$ and $\epsilon$.}
	\label{fig: z_rec}
\end{figure}

\begin{figure*}[t]
	\centering
	\includegraphics[width=\columnwidth]{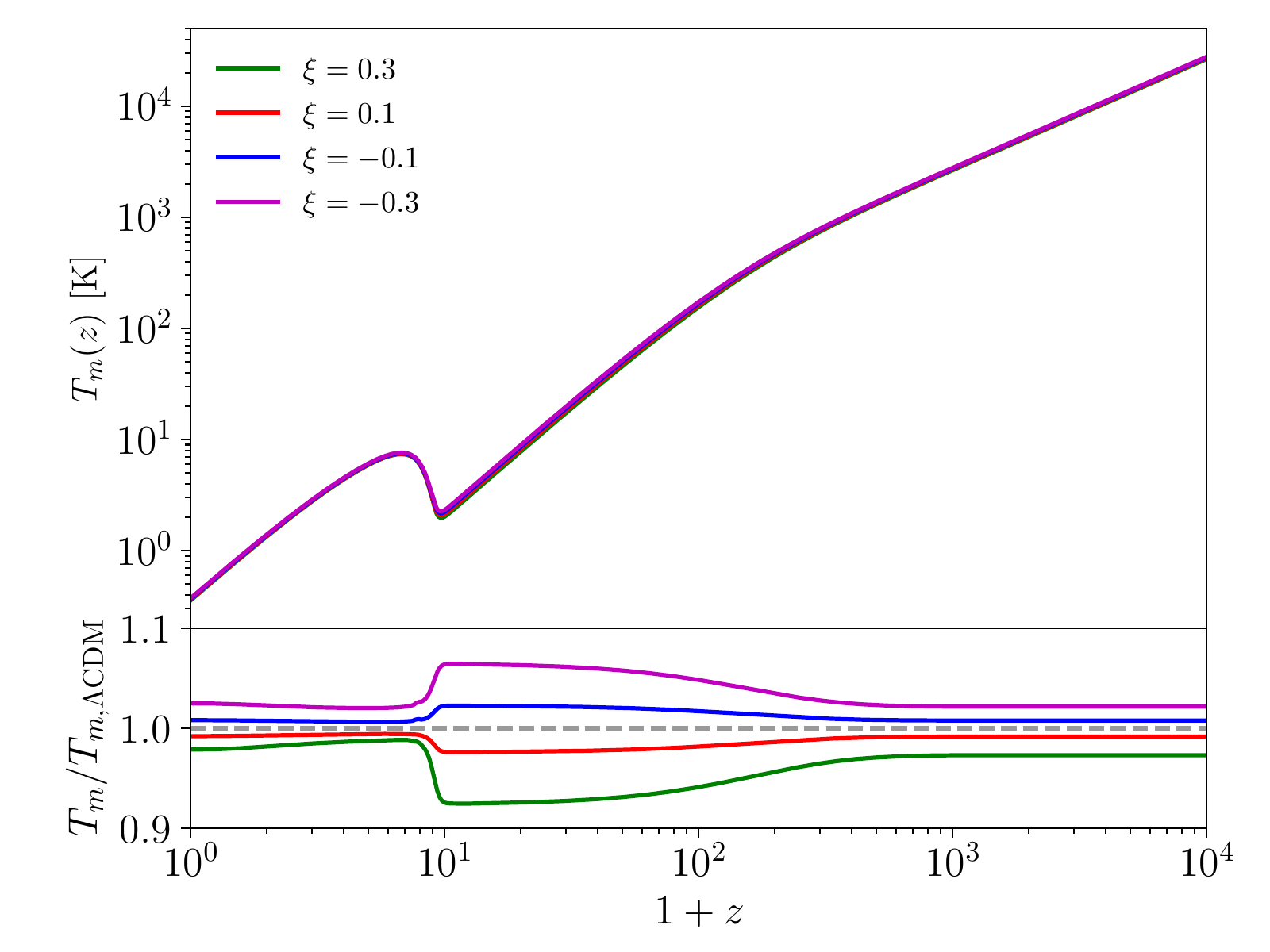}
	\includegraphics[width=\columnwidth]{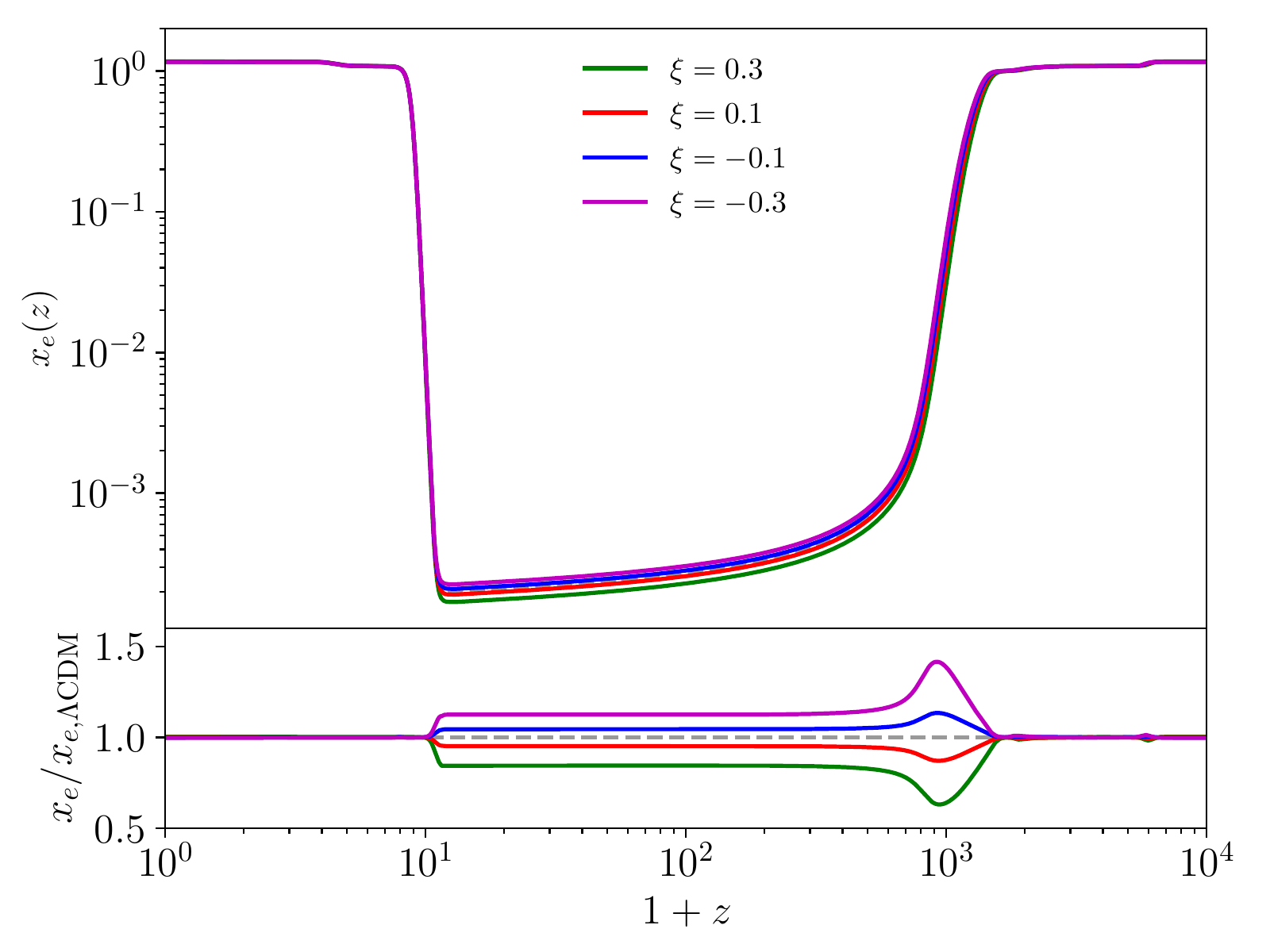}
	\caption{Effect of varying the free parameter of the iDE model considered within this work (in this case only $\xi$, since $\epsilon$ and $w_x$ are degenerate with it) on the evolution of the matter temperature $T_m$ (left panel) and of the free electron fraction $x_e$ (right panel). The lower panels show the same quantities as the upper panels, but normalized to the $\Lambda$CDM equivalent, which is reported as gray dashed line for reference.}
	\label{fig: evolution_Tm_xe}
\end{figure*}

Now that the main features of the modified temperature evolution have been outlined, we turn our attention to some of its most relevant consequences. First of all, it is important to remark the fact that the correspondence between photon temperature and redshift (${z=T_\gamma/T_{\gamma,0}-1}$ in the standard model) is modified. Indeed, assuming for sake of simplicity that at high $z$ the non-standard redshift dependence in Eq. \eqref{eq: T_g} can be neglected (which is true for $w_x$ close to its standard value, as clear from Fig.~\ref{fig: evolution_T}), one obtains that
\begin{align}
	z\simeq\left(\frac{T_\gamma}{T_{\gamma,0}}\right)\left(1+\frac{\epsilon \xi}{3w_{{\rm eff},x}-1}\frac{\rho_{x,0}}{\rho_{\gamma,0}}\right)^{-1/4}-1\,.
\end{align}
This means that in the iDE model considered here an event happening at a given temperature (like e.g., baryon-photon decoupling at around $T_\gamma=0.25$~eV) is shifted backward or forward in redshift, i.e., $z$ increases or decreases, with respect to the $\Lambda$CDM case depending on whether $\xi$ is positive or negative. The intuitive physical explanation is that when $\xi$ is positive (or negative) the energy density flows from the DE to the photons (or \textit{vice versa}), so that the photon temperature is lower (or higher) at earlier times (see Fig. \ref{fig: evolution_T}) and therefore it takes less (or more) time for the radiation to reach down to a given temperature. Concretely, considering for instance that in the $\Lambda$CDM model recombination\footnote{Here we adopt the same definition of recombination time as in CLASS, where it is taken to be the moment in time where the visibility function reaches its maximum value.} happens at $z_{d,\Lambda{\rm CDM}}=1088$, for $\xi=0.3,\,0.1,\,-0.1,\,-0.3$ (and $\epsilon=1\times10^{-4}$) we have that ${z_{d}=1115,\,1096,\,1081,\,1068}$. A graphical representation of this behavior is shown in Fig. \ref{fig: z_rec}, which also illustrates the role of $\epsilon$.

Another interesting effect of a modified thermal history such as the one considered here is that also the matter temperature $T_m$ changes (due to the tight coupling with the photons), and consequently also the free electron fraction $x_e$. Focusing first of all on the matter temperature, its evolution can be defined as  
\begin{align}\label{eq: dT_m}
	(1+z)\frac{\text{d}T_m}{\text{d}z}=\frac{\Gamma_C}{H}(T_m-T_\gamma)+2T_m\,,
\end{align}
where 
\begin{align}\label{eq: Gc}
	\Gamma_C\propto \frac{x_eT_\gamma^4}{1+f_{He}+x_e}
\end{align}
is the Compton scattering rate (see e.g., \cite{Seager1999New, Seager2000Exactly, Scott2009Matter} for the full equation and more details on the quantities involved). This implies that, as long as $\Gamma_C\gg H$ (i.e., when Compton scattering is still efficient) the Compton term dominates, so that photons and baryons can remain in thermal equilibrium with $T_m\simeq T_\gamma$. Therefore, for redshifts approximately above $10^3$ (where $x_e\simeq 1$) we expect that $T_m$ will present the very same offset as the one already observed for $T_\gamma$ in Fig. \ref{fig: evolution_T}. Then, below that threshold, after the two fluids decouple, the matter temperature starts to scale as $(1+z)^{2}$, but preserving the same temperature shift it had at the moment of decoupling. This means that we can expect in $T_m$ the same constant deviation with respect to $\Lambda$CDM over the whole history of the universe (which was not the case for $T_\gamma$ at low redshifts). 

On the other hand, the free electron fraction $x_e$ evolves according to 
\begin{align}\label{eq: dx_e}
	(1+z)\frac{\text{d}x_e}{\text{d}z}\propto x_e^2n_H\alpha_B-\beta_B(1-x_e)e^{-\nu_{2s}/T_m}
\end{align}
(see e.g., \cite{Seager1999New, Seager2000Exactly} for the complete equation and for more details on the quantities involved). Since at $z\geq 10^3$ electrons are fully ionized (i.e., $x_e\simeq 1$), in this case we see from Eq. \eqref{eq: dx_e} that there is no dependence of $x_e$ on deviations in $T_m$ from the standard picture, so that the evolution of $x_e$ is unaffected by our model. Then, at $z\simeq 10^3$ the electrons and protons recombine, and we expect two effects happening. The first one is that, as already hinted to before, since the matter temperature is increased or decreased, depending on the value of $\xi$, the redshift of recombination, and therefore the sharp drop in $x_e$, is delayed or anticipated. Then, after the free electron fraction reaches the plateau close to its minimum value the effect of an increased (decreased) matter temperature is that of decreasing (increasing) the exponential function in the second term of the RHS of Eq. \eqref{eq: dx_e}, thus increasing (decreasing) the overall amount of free electrons. Since the offset in $T_m$ is approximately constant, also the shift in $x_e$ during the plateau phase will be roughly constant. Finally, when reionization kicks in (at a redshift fixed by initial conditions) the standard picture is restored once again. In summary, we expect therefore a delayed (or anticipated) drop of $x_e$ followed by a constant offset between ${z\simeq 10-10^3}$. 

Importantly, note also that the proportionality of $\Gamma_C$ to $x_e$ highlighted in Eq. \eqref{eq: Gc} affects the evolution of $T_m$ between ${z\simeq 10-10^3}$, so that $T_m$ will have a slightly incremented deviation with respect to the standard scenario in this redshift interval. This will then affect the evolution of $x_e$ itself in the same way via the dependence on $T_m$ in Eq. \eqref{eq: dx_e}.

As an interesting remark, the impact of a modified $T_m$ such as the aforementioned one is conceptually similar to a scenario involving a constant energy injection, as it would be the case of DM annihilation with a constant annihilation efficiency (see e.g., \cite{Galli2009CMB} and in particular Fig.~1 therein). This can be understood more formally focusing on e.g., Eq. (9) of \cite{Galli2009CMB} and substituting their extra term coming form the DM annihilation with the constant temperature offset $\Delta T_\gamma$ that we get from the modified $T_\gamma$ evolution. In other words, expanding the Compton scattering term in Eq. \eqref{eq: dT_m} with $(T_m-T_{\gamma,\Lambda\text{CDM}}-\Delta T_\gamma)$ one would obtain a contribution comparable to the extra term in Eq. (9) of \cite{Galli2009CMB}, assuming that the DM annihilation rate is constant in time. 

This whole discussion can be quantitatively checked against the concrete examples of how $\xi$ affects $T_m$ and $x_e$ over the history of the universe that we propose in Fig.~\ref{fig: evolution_Tm_xe} (where we follow the same color-coding as in Fig.~\ref{fig: evolution_T}). As expected, in the left (bottom) panel one can see that at early times the deviation in $T_m$ from the $\Lambda$CDM case is constant and exactly equal to the offset already shown in Fig. \ref{fig: evolution_T} for $T_\gamma$. However, after recombination the additional proportionality to $x_e$ kicks in and the deviations are enhanced, only to reduce again after reionization begins. Similarly, also for $x_e$ there are no appreciable deviations before recombination, as expected. Then, for positive (negative) values of $\xi$ we have that the vertical drop at around $z\simeq10^3$ happens earlier (later) than in the standard model, leading to the large deviation in the bottom panel. However, once the plateau is reached one has the predicted nearly constant offset (the aforementioned increment expected due to the enhancement in $T_m$ cannot be seen in the plot but it is there).

\subsection{Perturbations}\label{sec: math_pt}
The derivation of the perturbation equations for this particular setup has been, to our knowledge, proposed for the first time in App. \ref{sec: app per}. For simplicity, we work in the synchronous gauge and refer to \cite{Ma1995Cosmological} for more details on the notation and the standard contributions. 

The resulting set of equations for the DM reads
\begin{align}
	& \dot{\delta}_c =-\frac{\dot{h}}{2}+(1-\epsilon)\xi\frac{\rho_x}{\rho_c}\left[\frac{\dot{h}}{6}+H(\delta_x-\delta_c)\right]\,, \\ & \dot{\theta}_c=0 \label{eq: dot theta_c}\,,
\end{align}
where Eq. \eqref{eq: dot theta_c} follows from our gauge choice and the assumption that no momentum is transferred between the fluids. Similarly, for the photons we have that 
\begin{align}
	\nonumber & \dot{\delta}_\gamma= [\dots] +\epsilon \xi\frac{\rho_x}{\rho_\gamma}\left[\frac{\dot{h}}{6}+H(\delta_x-\delta_\gamma)\right]\,,
\end{align}
where the $[\dots]$ refers to the standard terms already accounted for in Eq. (58) of \cite{Ma1995Cosmological}, while the equations for $\dot{\theta}_\gamma$ and the momentum averaged phase-space density perturbation remain unchanged with respect to Eq. (58) of~\cite{Ma1995Cosmological}. Finally, the perturbation equations of the DE fluid read
\begin{align}
	\nonumber \dot{\delta}_x = & -(1+w_x)\left(\theta_x+ \frac{\dot{h}}{2}\right) -\xi\frac{\dot{h}}{6}\\ & -3H\left(c_{s,x}^2-w_x\right)\left[\delta_x+\frac{H\theta_x}{k^2}(3(1+w_x)+\xi)\right]\,, \\
	\nonumber \dot{\theta}_x= &  2H\theta_x\left[1+\frac{\xi}{1+w_x}\left(1-(1-\epsilon)\frac{\theta_c}{2\theta_x} -\frac{4\epsilon}{3}\frac{\theta_\gamma}{2\theta_x}\right)\right] \\ 
	& +\frac{c_{s,x}^2 k^2\delta_x}{1+w_x} \label{eq: dot theta_x 2}\,.
\end{align}
Suitable initial conditions for these equations are also discussed in the appendix.
\begin{figure*}
	\centering
	\includegraphics[width=\columnwidth]{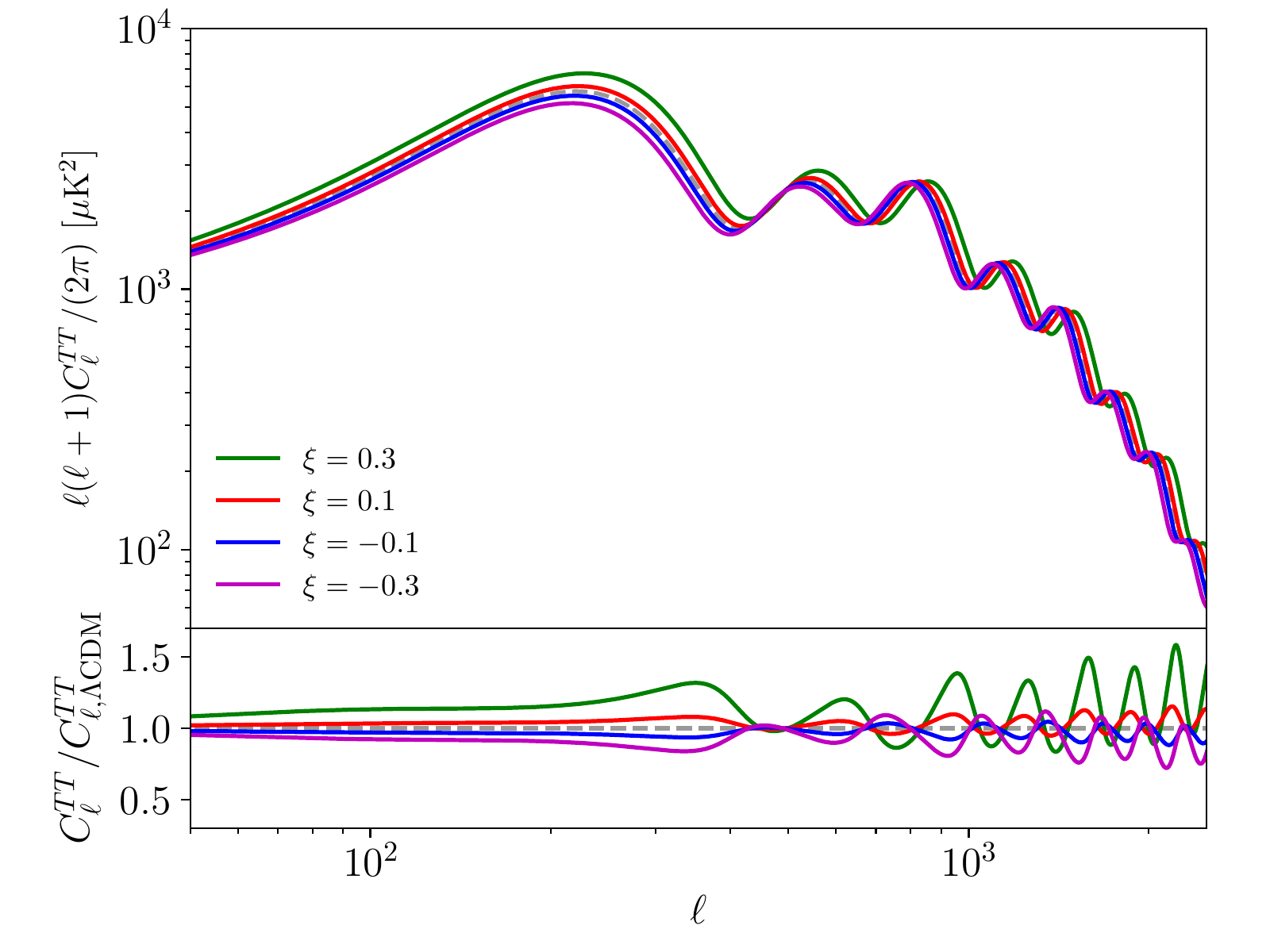}
	\includegraphics[width=\columnwidth]{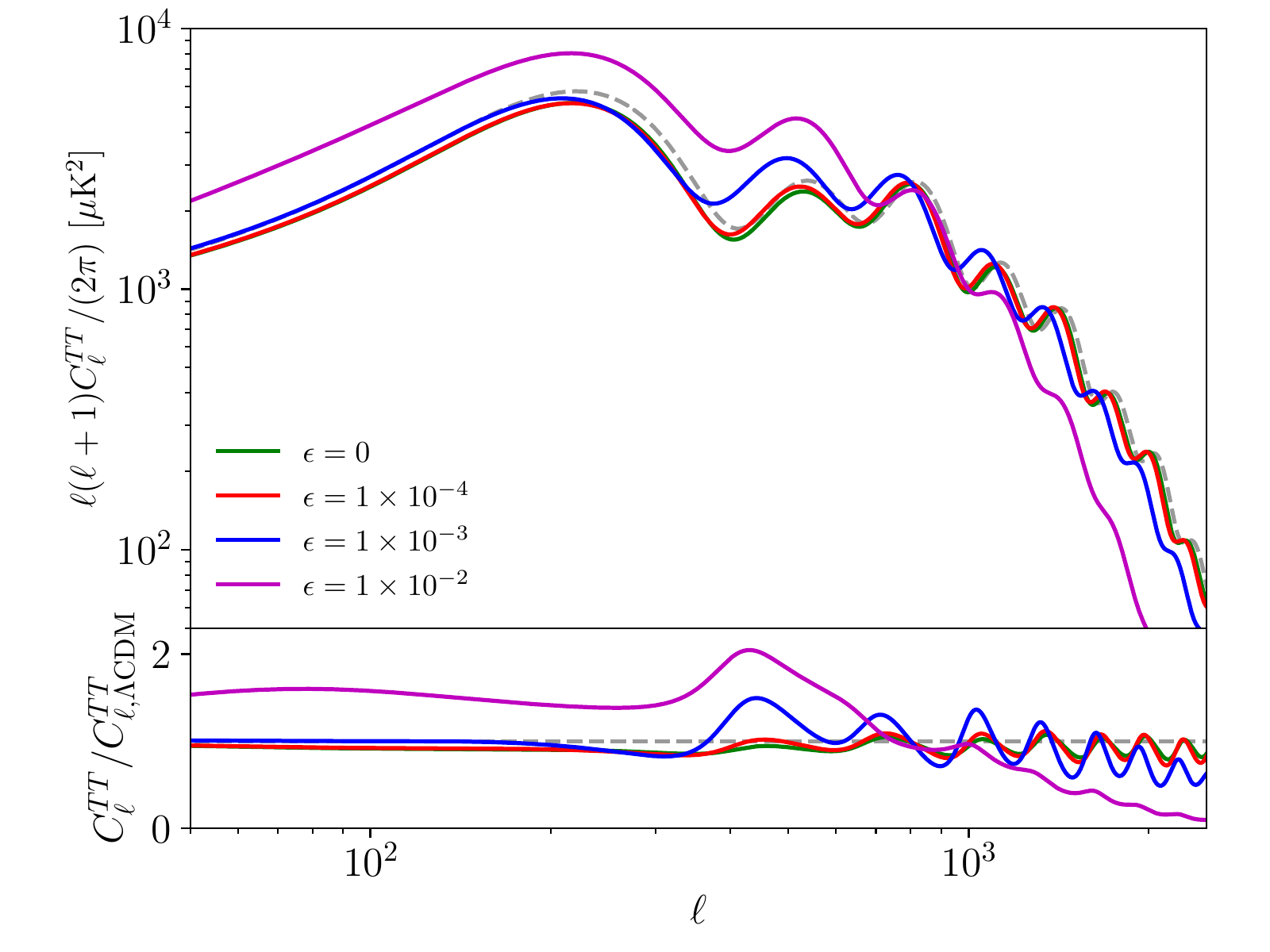}
	\caption{Effect of varying the free parameters of the model ($\xi$, left panel with $\epsilon=1\times10^{-4}$, and $\epsilon$, right panel with $\xi=-0.3$) on the CMB temperature power spectrum. As in the previous figures, the $\Lambda$CDM case is reported as gray dashed line for reference.}
	\label{fig: evolution_TTm}
\end{figure*}

As a remark, it is important to confirm that for $\epsilon=0$ we recover the same well-known equations for the interacting DM-DE case, which are for instance summarized in Eqs. (2.7)-(2.10) of \cite{Lucca2020Shedding}. Therefore, the derivation presented in App. \ref{sec: app per} also answers the concerns raised in Sec. 2 of \cite{Lucca2020Shedding} regarding the presence of multiple sets of perturbation equations for that particular model.

Finally, as firstly noted in \cite{Gavela2009Dark} for the $\epsilon=0$ case, great care has to be taken in the choice of the priors on the free parameters of the model to avoid early time instabilities in the DE perturbation equations. In particular, following Secs. 2.2 and 3.4 of the reference, it turns out that the instabilities are avoided if $\xi$ and $w_x+1$ are of opposite sign, i.e., if the so-called doom factor mentioned there is negative. Since both the doom factor and the form of Eq.~(42) in the reference are independent of whether the coupling to photons is activated or not\footnote{The reason for this is that Eq. (18), which has been used to derive Eq. (42), separates the DE dependent terms (with coefficients $A_e$ and $B_e$) from those depending on any other fluid (in our case both the DM and the photons). Then, in Eq. (42) only the former appear, meaning that the latter are subdominant, regardless of the kind of considered fluids.}, we can extend their conclusions to our scenario as well.

\section{Impact on the observables}\label{sec: obs}
Now that the full set of cosmological equations has been derived in the context of the iDE model considered within this work, we can turn our attention to the impact that  the model's parameters have on observables such as the CMB anisotropy power spectra and the matter power spectrum.

\subsection{CMB anisotropy power spectra}\label{sec: obs_TT}
First of all, we focus on the temperature power spectrum\footnote{Complementary discussions for the specific case with only DM-DE interactions ($\epsilon=0$) can be found in e.g., \cite{Murgia2016Constraints}.} and in Fig. \ref{fig: evolution_TTm} we show the effect of varying $\xi$ (left panel, with $\epsilon=1\times10^{-4}$) and $\epsilon$ (right panel, with $m=-0.3$). In the left panel of the figure one can recognize two effects: depending on whether $\xi$ is positive or negative we have 1) an enhancement or suppression of the first two peaks and 2) an overall shift of the spectrum to the right or to the left. The first effect can be explained by the modified redshift of matter-radiation equality $z_{\rm eq}$, while the second appears due to the changes in the angular scale of the sound horizon $\theta_s$.

In particular, let us assume that $\xi$ is negative (and $\epsilon$ is small), so that $z_{\rm eq}$ is higher than in the $\Lambda$CDM case (i.e., matter-radiation equality happens earlier, see Fig.~\ref{fig: z_eq}). This would then imply that the early integrated Sachs-Wolfe (ISW) effect, which builds up at scales crossing the sound horizon during RD, is suppressed. As a consequence, since the ISW effect mainly contributes to the amplitude of the first peak of the temperature power spectrum, the latter is also reduced. Furthermore, because of the higher $z_{\rm eq}$ value, there must be some scales that enter the Hubble radius during MD instead of during RD, as they would in the standard case. These would then not feel the gravitational boost that takes place during RD after Hubble crossing and, since these scales are typically those that contribute to the second peak of the spectrum, also the second peak is suppressed.\footnote{For a very exhaustive pedagogical description of these effects see e.g., \cite{Lesgourgues:2013qba} with particular emphasis on Sec. 2.5 therein.} The sum of these two effects can be clearly observed in the first two peaks of the spectrum in the right panel of Fig. \ref{fig: evolution_TTm}. Of course, the opposite effect is to be expected for $\xi$ positive (as long as $\epsilon$ is small).

Furthermore, the angular scale of the sound horizon $\theta_s$ is well-known to set the angular scale of the first peak of the spectrum $\ell_s=\pi/\theta_s$ and can be shown to be related to the position of the other peaks in the spectrum $\ell_N$ via the relation (see e.g., \cite{Lesgourgues:2013qba})
\begin{align}\label{eq: theta_N}
	\frac{\theta_s}{N}=\frac{\pi}{\ell_N}\,,
\end{align}
where $N=2,3,4, ...$\,. This implies that when $\theta_s$ varies every peak of the spectrum is shifted to higher or lower $\ell$ values. On the other hand, the value of $\theta_s$ can be seen as the ratio of the sound horizon $r_s$ evaluated at decoupling ($z=z_d$) and the last-scattering distance $r_a$, i.e.,
\begin{align}
	\theta_s=\frac{r_s(z_d)}{r_a(z_d)}\,.
\end{align}
The former can be defined as
\begin{align}\label{eq: rs}
	r_s(z_d)=\int_{z_d}^{\infty}\frac{c_s(z)\text{d}z}{H(z)}\,,
\end{align}
and it is only affected by the evolution of the universe prior to recombination, while the latter is determined by the history of the universe after decoupling according to the relation
\begin{align}
	r_a(z_d)=\int_{0}^{z_d}\frac{\text{d}z}{H(z)}\,.
\end{align}
In Eq. \eqref{eq: rs}, $c_s(z)$ is the sound speed of the baryon-photon plasma defined in Eq. \eqref{eq: cs}.
\begin{figure*}
	\centering
	\includegraphics[width=\columnwidth]{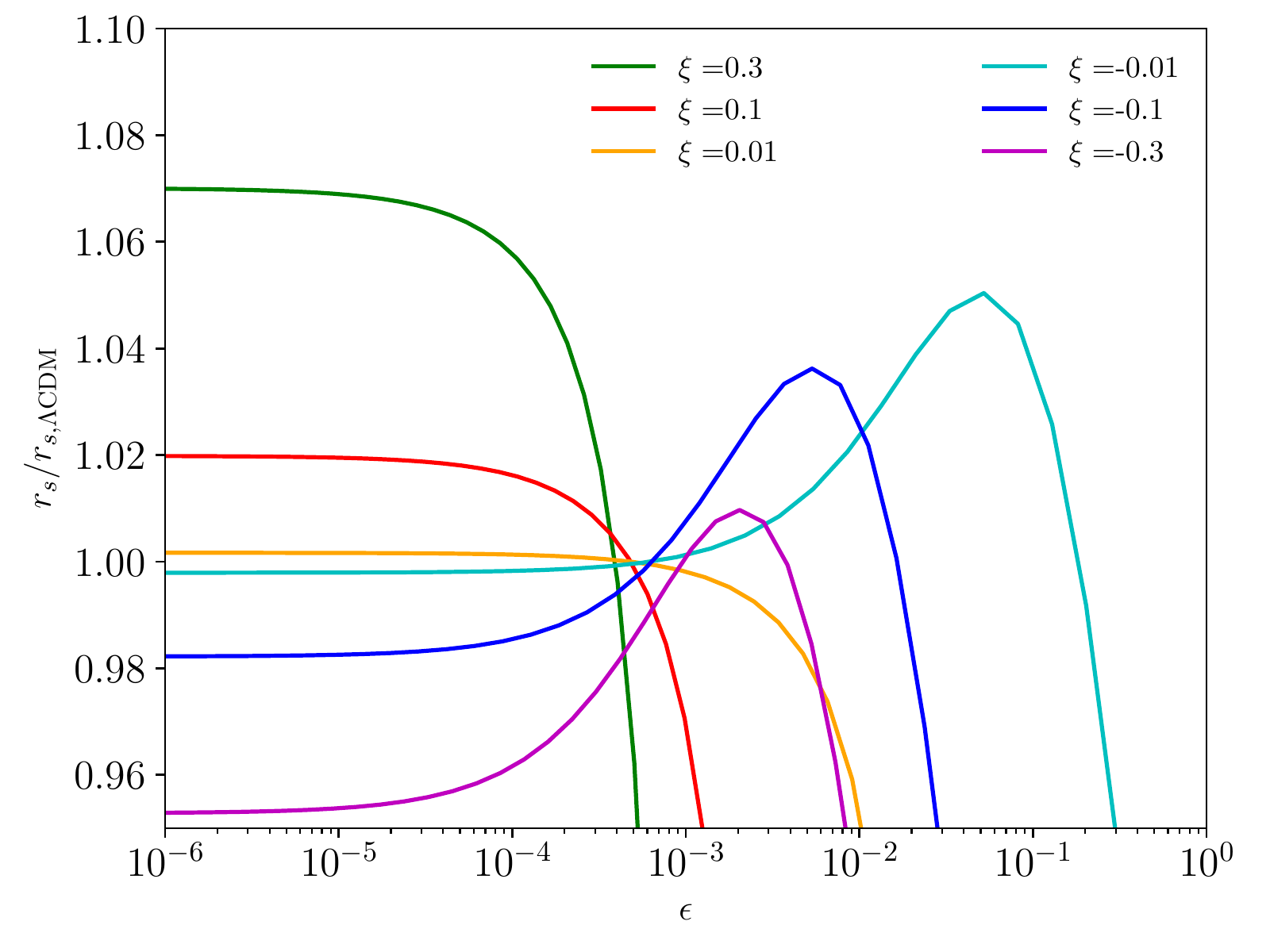}
	\includegraphics[width=\columnwidth]{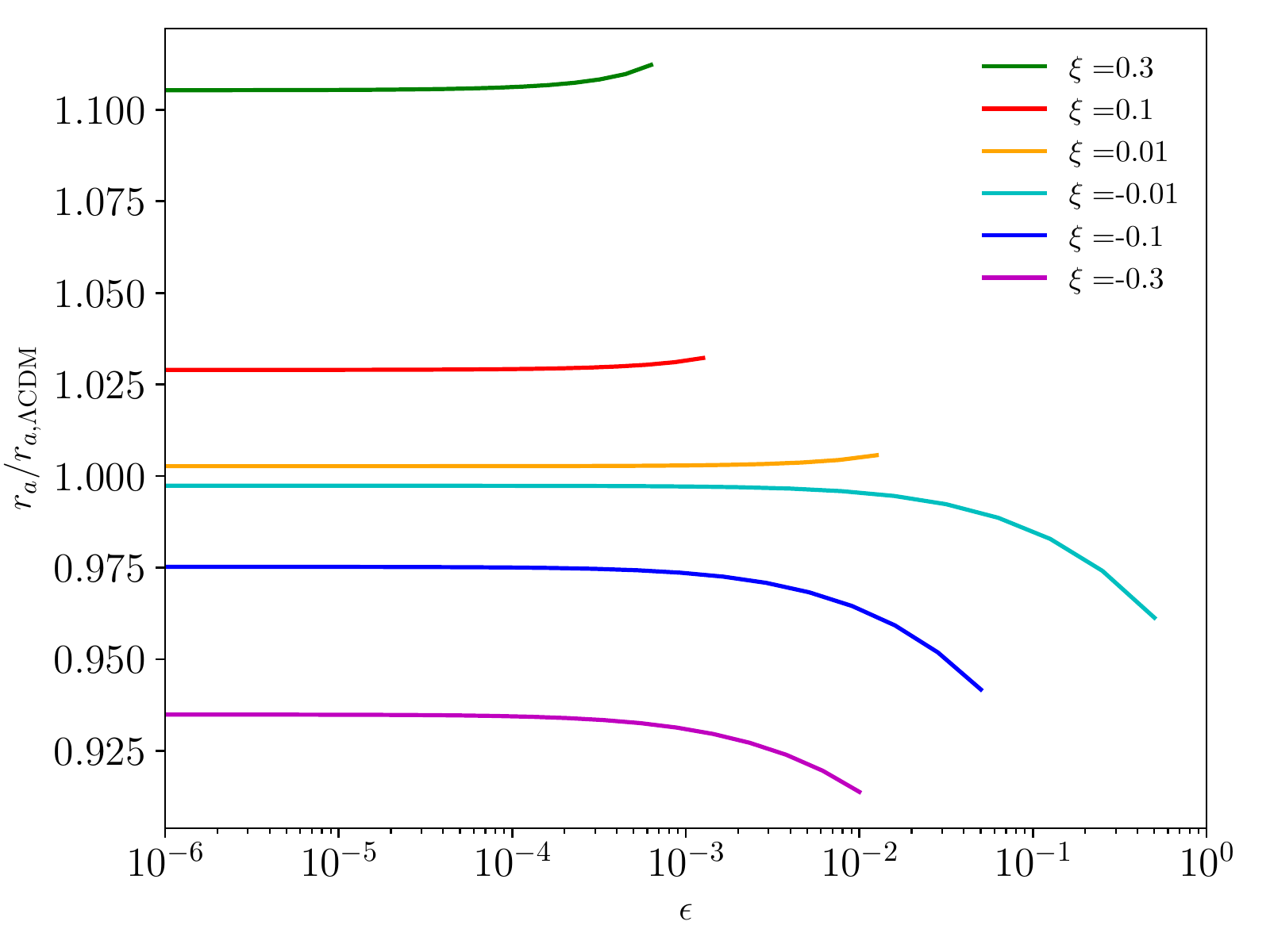}
	\caption{Dependence of the sound horizon $r_s$ (left panel, normalized to the $\Lambda$CDM value $r_{s,\Lambda{\rm CDM}}=144.7$ Mpc) and the last-scattering distance $r_a$ (right panel, normalized to the $\Lambda$CDM value $r_{a,\Lambda{\rm CDM}}=1.391\times10^5$ Mpc) on the free parameters of the model $\xi$ and $\epsilon$. In the right panel the lines are interrupted whenever an assumption required by CLASS is violated.}
	\label{fig: rs}
\end{figure*}

From the discussion outlined in Sec. \ref{sec: math}, it is then clear that a given variation in $\xi$ affects both $r_s$ and $r_a$ in multiple competing ways. As in the example above, let us assume that $\xi$ is negative. Then, first of all we have that $z_d$ decreases (see Sec. \ref{sec: math_th} and in particular Fig. \ref{fig: z_rec}), so that the integration intervals change and therefore $r_s$ increases while $r_a$ decreases. Secondly, from Eq.~\eqref{eq: cs} it becomes clear that in the scenario considered so far $c_s$ is increased with respect to the standard value as $\rho_\gamma$ grows while $\rho_b$ is unaffected. This also leads to an higher value of $r_s$. However, a negative value of $\xi$ would increase $H(z)$ both prior and after recombination (as shown in the right top panel of Fig. \ref{fig: evolution_O}), so that both $r_s$ and $r_a$ decrease. In the case of $r_s$, while the first two effects lead to an higher value, the third effect is by far the most dominant (for $\epsilon=1\times10^{-4}$) and therefore we expect a final reduction of the value of $r_s$. On the other hand, in the case of $r_a$, the first and the third effect sum up leading to a lower value, and since $H$ is also affected after recombination much more than prior to that, $r_a$ decreases more strongly than $r_s$. As a consequence, $\theta_s$ is increased and we see from Eq.~\eqref{eq: theta_N} that all the peaks (positioned at $\ell_N$) shift to lower values, as also observed in the right panel of Fig.~\ref{fig: evolution_TTm}. As long as $\epsilon$ is small, the opposite behavior applies for positive values of~$\xi$.

Moving on to the role of $\epsilon$, one can recognize from the right panel of Fig. \ref{fig: evolution_TTm} a number of effects. Firstly, the overall horizontal shift of the peaks is enhanced. The reason for this can be understood following the behavior of $r_s$ when $\epsilon$ is increased. Indeed, as already mentioned before (see the middle right panel of Fig. \ref{fig: evolution_O} as well as Fig. \ref{fig: z_rec}), for negative (positive) values of $\xi$ of the order of $-0.3$ ($+0.3$) when $\epsilon$ is small (below $1\times10^{-4}$) $z_d$ and $c_s$ are almost unchanged, while $H$ is already significantly increased (decreased), in particular by the contribution of the DM during MD prior to recombination. Therefore we expect $r_s$ to decrease (increase) proportionally only to $\xi$. Moreover, for values of $\epsilon$ of the order of $1\times10^{-3}$, $z_d$ diverges much faster than $H$, so that $r_s$ increases (decreases) very rapidly. However, when $\xi$ is negative $\epsilon$ can be increased to arbitrarily high values so that the Hubble rate diverges much faster than $z_d$ and $r_s$ decreases again. On the other hand, according to the conditions derived in Sec. \ref{sec: math_bg} from \text{Eqs. \eqref{eq: rho_c} and \eqref{eq: rho_g}}, when $\xi$ is positive $\epsilon$ cannot be arbitrarily high, so that very high values of $H$ cannot be reached and $r_s$ is always decreasing. Furthermore, since a variation of $\epsilon$ mainly affects the RD era, it only impacts $r_a$ via the eventual changes in $z_d$, which are however subdominant with respect to the modifications caused by the non-standard evolution of $H$ until $z_d$ strongly diverges. For this reason we expect $r_a$ to be almost independent of $\epsilon$ up to $\epsilon\simeq10^{-3}$, and then to decrease or increase depending on whether $\xi$ is negative or positive. A general dependence of $r_s$ (left panel) and $r_a$ (right panel) on $\xi$ and $\epsilon$ is shown in Fig. \ref{fig: rs}. There, all of the aforementioned behaviors are confirmed. 

Therefore, to summarize, for $\xi=-0.3$ as in the right panel of Fig.~\ref{fig: evolution_TTm}, when $\epsilon$ is increased up to $1\times10^{-2}$ (which corresponds to the threshold at which CLASS breaks down) $r_s$ is almost always lower than in the $\Lambda$CDM case, but not as much as $r_a$. Therefore, the more $\epsilon$ is increased, the lower the value of $\theta_s$ becomes and consequently all peaks shift more and more to the left, as observed in the right panel of Fig.~\ref{fig: evolution_TTm}.

Then, as clear from Fig. \ref{fig: z_eq}, increasing the value of $\epsilon$ leads to lowering of $z_{\rm eq}$. Therefore, following the aforementioned discussion involving the early ISW effect and the gravitational boosting, we expect for negative values of $\xi$ an increasing enhancement of the first two peaks the higher the value of $\epsilon$ becomes. This is precisely what happens in the right panel of Fig. \ref{fig: evolution_TTm}.

Another effect clearly present in the right panel of Fig.~\ref{fig: evolution_TTm} is that the higher the value of $\epsilon$ is, the stronger the damping at large scales becomes. To explain this behavior we can proceed analogously as before and define a damping scale $r_D$, which has a similar dependence on $\xi$ and $\epsilon$ as $r_s$ (see e.g., \cite{Hu:1996mn, Lesgourgues:2013qba} for a definition and more in-depth discussions). We can consequently also define the associated angular scale
\begin{align}
	\theta_D=\frac{r_D}{r_a}=\frac{\pi}{\ell_D}\,,
\end{align}
where $\ell_D$ represents the multipole at which the exponential damping begins. Then, as for $\theta_s$, also $\theta_D$ decreases with increasing $\epsilon$ (for $\xi$ negative), which implies that $\ell_D$ decreases and the damping is stronger. Precisely this is what we see in the right panel of Fig.~\ref{fig: evolution_TTm} (and also in the left panel, although in a less visible way).

\subsection{Matter power spectrum}\label{sec: obs_Pk}
\begin{figure}[t]
	\centering
	\includegraphics[width=0.95\columnwidth]{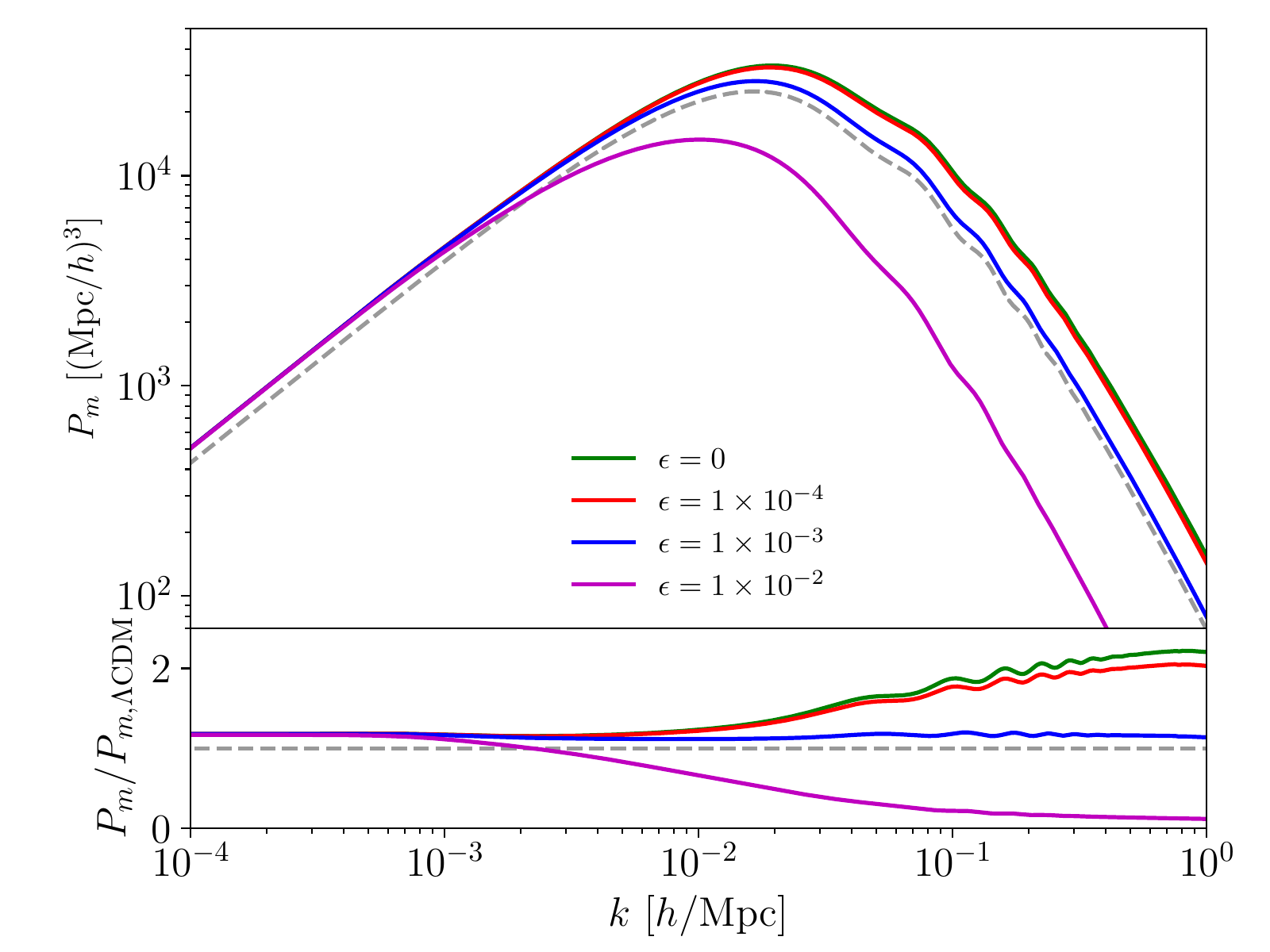}
	\caption{Same as the right panel of Fig. \ref{fig: evolution_TTm} but for the matter power spectrum.}
	\label{fig: evolution_Pk}
\end{figure}

We move on now to the impact of $\xi$ and $\epsilon$ on the matter power spectrum. As in the previous section, a graphical representation of the effects that we will discuss below can be seen in Fig. \ref{fig: evolution_Pk} for the case of $\epsilon$, while an analogous depiction focused on the role of $\xi$ can be found in Fig.~1 of \cite{Lucca:2021dxo} (where $\epsilon=0$ was assumed, but is still accurate as long as $\epsilon$ is smaller than $\simeq1\times10^{-4}$, as evident from Fig.~\ref{fig: evolution_Pk}).

We start by focusing on the role of $\xi$ by briefly summarizing the discussion already presented in \cite{Lucca:2021dxo}. As pointed out in the reference, $\xi$ impacts the matter power spectrum mainly in two ways. On the one hand, since both $\omega_m$ and $z_{\rm eq}$ depend on $\xi$, so does $k_{\rm eq}$ (see e.g., Eq.~6 of \cite{Lucca:2021dxo}) and hence the position of the peak of the spectrum (shifting to the left/right depending on whether $\xi$ is positive/negative). On the other hand, for $\xi$ positive or negative one observes an overall suppression or enhancement of the spectrum, in particular at scales higher than~$k_{\rm eq}$.

Next, we consider the impact of $\epsilon$ graphically displayed in Fig. \ref{fig: evolution_Pk} (where we assume $\xi=-0.3$ as before). From the figure it is clear that, as one would expect, large scales (built up during DE and matter domination) are not affected by $\epsilon$. However, the larger the value of $\epsilon$ is, the smaller the values of $z_{\rm eq}$ (see Fig. \ref{fig: z_eq}) and correspondingly of $k_{\rm eq}$ (see Eq. 6 of \cite{Lucca:2021dxo}) become, so that the peak of the spectrum shifts more and more to the left. Furthermore, because of the same effect that $\epsilon$ has on $k_{\rm eq}$, also the amplitude of the spectrum roughly above $k_{\rm eq}$ (mainly proportional to $k_{\rm eq}$, $(1-\omega_b/\omega_{m})$ and $\eta_0$, see e.g., \cite{Lucca:2021dxo}) decreases for increasing values of $\epsilon$ (and $\xi$ kept constant). Interestingly, this means that it is possible for $\epsilon$ to compensate the large deviations induced by $\xi$ at large scales, as shown by the blue curve in the figure (with $\epsilon=1\times10^{-3}$).

\section{Method and cosmological probes}\label{sec: meth}
The mathematical setup described in Sec. \ref{sec: math} has been implemented in the Boltzmann solver CLASS \cite{Lesgourgues2011Cosmic, Blas2011Cosmic} (v2.9) and the code has been made publicly available\footnote{\url{https://github.com/luccamatteo/class_iDMDEg}}. This modified version of CLASS has then been used in combination with the parameter inference code MontePython \cite{Audren2013Conservative, Brinckmann2018MontePython} to perform a series of Markov Chain Monte Carlo (MCMC) scans on the free parameters of the model introduced in Sec. \ref{sec: math}. 

Overall, we will consider two new possible scenarios. In the first one we only consider DE-photon interactions (referred to henceforth as iDE$\gamma$ model), and fix therefore $\epsilon=1$. In this case we have a 6+1 extension of the standard $\Lambda$CDM model with
\begin{align}\label{eq: params set e1}
	\{H_0,\, \omega_b,\, \omega_{\rm cdm},\, n_s,\, \ln(10^{10}A_s),\, \tau_{\rm reio} \} + \xi\,.
\end{align}
Secondly, we generalize this situation to account for both interactions, leading to a 6+2 extension with
\begin{align}\label{eq: params set efree}
\{H_0,\, \omega_b,\, \omega_{\rm cdm},\, n_s,\, \ln(10^{10}A_s),\, \tau_{\rm reio} \} + \xi,\,\epsilon\,.
\end{align}
We will refer to this case as iDMDE$\gamma$ model. Following the discussion had in Sec. \ref{sec: math_pt}, for each parameter set we have to distinguish between the two scenarios where $\xi$ is positive or negative, thereby fixing\footnote{In the context of this paper we will not consider the case where the DE EOS parameter is free to vary since it does not come with any clear degeneracy with $\xi$ or $\epsilon$ (see Sec. \ref{sec: math}) and is known to be strongly constrained by the data set combinations discussed below. We leave a more detailed analysis in this direction for future work.} the value of $w_x$ respectively to $-1.001$ and $-0.999$, as often done in the literature. For all standard $\Lambda$CDM parameters we assume flat, unbounded priors, while for $\xi$ and $\epsilon$ we assume flat priors of the form $\xi>0$ (or $\xi<0$ depending on the chosen model) and constrained in the range $\epsilon\in[0,1]$, respectively.

For the second limiting case, the iDMDE model with $\epsilon=0$, we refer the reader to \cite{Lucca2020Shedding} and \cite{Lucca:2021dxo} (for $\xi$ negative and positive, respectively), which conduct an analysis very similar in method and spirit to one proposed here (see also e.g., \cite{DiValentino2017Interacting, DiValentino2019Interacting} for additional discussions).

For the choice of cosmological data sets employed to test the validity of the iDE models described above we follow \cite{Lucca:2021dxo}. In particular, we adopt the same \textit{baseline} combination of probes which relies on data from \textit{Planck} 2018~\cite{Aghanim2018PlanckVI} (temperature, polarization and lensing likelihoods), the 6dF Galaxy Survey (6dFGS) \cite{Beutler2011Galaxy}, the Sloan Digital Sky Survey (SDSS), the Baryon Oscillation Spectroscopic Survey (BOSS) \cite{Ross2014Clustering, Alam2016Clustering} and the SNIa Pantheon catalog~\cite{Scolnic2017Complete}. As in \cite{Lucca:2021dxo}, here we also refrain from including late-time priors on the $H_0$ parameter following \cite{Benevento:2020fev, Camarena:2021jlr, Efstathiou:2021ocp}. For the iDMDE$\gamma$ scenario, which as we will see is able to restore the concordance between early-time inference and late-time measurements of the $S_8$ value considering only the \textit{baseline} data sets, we will also include data from the combination of KiDS and the VISTA Kilo-Degree Infrared Galaxy Survey (VIKING)~\cite{Hildebrandt:2018yau}, henceforth referred to as KV450, as well as from DES~\cite{Abbott:2017wau, Troxel:2017xyo}. For the former we employ the publicly available MontePython likelihood\footnote{As a note, since for this particular iDE model, to our knowledge, no dedicated study of the evolution of the matter power spectrum at non-linear scales has been performed (and hence no numerical implementation in CLASS is possible), we employ only the linear scales covered by the likelihood. The distinction between linear and non-linear scales is defined automatically by the likelihood.} \cite{Hildebrandt:2018yau}, while we parameterize the latter as a Gaussian prior on $S_8$ of the form $S_8=0.782\pm0.027$ \cite{Troxel:2017xyo} (see~\cite{Lucca:2021dxo} for additional details and related discussions).

Finally, we test the statistical significance of our results using three different parameters: the simple $\Delta\chi^2$, the significance $\sigma$ and the Bayes ratio $B_{\rm iDE,\Lambda}$, for whose calculation we make use of the \texttt{MCEvidence} code~\cite{Heavens:2017afc} (referring to e.g., \cite{Gomez-Valent:2020mqn} for the method and the theoretical aspects). For the characterization of the latter we will make use of the Jeffreys' scale \cite{doi:10.1080/01621459.1995.10476572}, according to which a negative (positive) value of $\ln B_{\rm iDE\gamma,\Lambda}$ would point to a preference for the $\Lambda$CDM (iDE) model. Quantitatively, preferences in the ranges between $0-5$, $5-10$ and $10-15$ will be referred to as \textit{mild}, \textit{substantial} and \textit{strong}, respectively.

\section{Cosmological constraints}\label{sec: res}
In this section we will discuss the numerical results obtained by confronting the iDE model introduced in Sec.~\ref{sec: math} against the several cosmological probes discussed in Sec.~\ref{sec: meth}. As mention in the latter section, we will investigate two different scenarios, the iDE$\gamma$ and iDMDE$\gamma$. 

We focus first of all on the iDE$\gamma$ case. The resulting constraints on the parameters mostly affected by the DE interactions are reported in Tab. \ref{tab: MCMC_eps} for both possible priors on $\xi$ and employing the full \textit{baseline} data set combination. In the table we also show both the derived parameter $\sigma_8$ and its common reformulation $S_8=\sigma_8(\Omega_m/0.3)^{0.5}$ to highlight the eventual model's ability to address the related tension. To our knowledge, these are the first constraints on this interacting model present in the literature from which we obtain for the coupling parameter~that
\begin{align}
	-1.4\times10^{-5}<\xi<1.1\times10^{-5}\,.
\end{align}
Additionally, in the bottom rows of the table we also list the three statistical parameters we will use to asses the significance of a given iDE scenario with respect to the $\Lambda$CDM model. 
\begin{table}[t]
	\centering
	\begin{tabular}{|c|c|c|}
		\hline\rule{0pt}{3.0ex} 
		Parameter & $\xi>$1 & $\xi<$1 \\[0.1 cm]
		\hline
		\rule{0pt}{3.0ex}
		$H_0 \, [\text{km/(s Mpc)}]$ & $68.49_{-0.56}^{+0.49}$ & $67.64_{-0.50}^{+0.56}$ \\[0.1 cm]
		$100\,\omega_{\rm b}$        & $2.214_{-0.019}^{+0.028}$ & $2.269_{-0.031}^{+0.020}$ \\[0.1 cm]
		$\omega_{\rm cdm}$           & $0.11943_{-0.00095}^{+0.00094}$ & $0.11971_{-0.00094}^{+0.00096}$ \\[0.1 cm]
		$\xi$                          & $<1.1\times10^{-5}$ & $>-1.4\times10^{-5}$ \\[0.1 cm]
		$\sigma_8$                   & $0.8296_{-0.0089}^{+0.0072}$ & $0.8143_{-0.0074}^{+0.0092}$ \\[0.1 cm]
		$S_8=\sigma_8(\Omega_m/0.3)^{0.5}$ & $0.829\pm0.011$ & $0.832\pm0.011$ \\[0.1 cm]
		\hline
		\rule{0pt}{3.0ex}
		$\Delta\chi^2$               & $-0.6$ & $-0.4$ \\[0.1 cm]
		$\sigma$             		 & $0.8$ & $0.6$ \\[0.1 cm]
		$\ln B_{\rm iDE\gamma,\Lambda}$          & $-11.6$ & $-11.5$ \\[0.1 cm]
		\hline
	\end{tabular}
	\caption{Mean and 68\% C.L. (the lower/upper bounds are given at the 95\% C.L.) of the parameters most significantly affected by the DE-photon interactions ($\epsilon=1$ case) for different choices of $\xi$ and the \textit{baseline} data set combination, together with the corresponding $\Delta\chi^2$, $\sigma$, and $\ln B_{\rm iDE\gamma,\Lambda}$ values.}	
	\label{tab: MCMC_eps}
\end{table}

From the table, it is clear that the model does not significantly alleviate any of the two cosmological tensions discussed above, with mean values and uncertainties in overall good agreement with the \textit{Planck}+BAO values quoted in Tabs. 1 and 2 of \cite{Aghanim2018PlanckVI} for the $\Lambda$CDM scenario. The same is true regardless of the prior on $\xi$. The reason for this might rely on the fact that this type of interaction impacts the expansion and thermal history of the universe in a too widespread and deep way, imprinting features in the observables that cannot be accommodated by the data nor reabsorbed by possible degeneracies with the other cosmological parameters.

Nevertheless, as hoped, in the case where $\xi$ is positive (which is able to solve the $S_8$ tension if $\epsilon=0$ \cite{Lucca:2021dxo}) the $H_0$ value is trending in the right direction, although not yet to an extent where it can restore the concordance between early- and late-time observations, while the $S_8$ value is almost unaffected with respect to the $\Lambda$CDM model. This leaves the window open to the possibility that degeneracies between $\xi$ and $\epsilon$ (as well as with the other $\Lambda$CDM parameters) would allow for higher values of $\epsilon$ than $\mathcal{O}(10^{-5})$, enhancing the \tquote{positive} aspects of the iDM$\gamma$ model in the more general context of the iDMDE$\gamma$ scenario, which should already be able to address the $S_8$ tension due to the DM-DE interactions~\cite{Lucca:2021dxo}.
\begin{table}[t]
	\centering
	\begin{tabular}{|c|c|c|}
		\hline\rule{0pt}{3.0ex} 
		Parameter & \textit{baseline} & \textit{baseline}+KV450+DES \\[0.1 cm]
		\hline
		\rule{0pt}{3.0ex}
		$H_0 \, [\text{km/(s Mpc)}]$ & $68.21_{-0.55}^{+0.58}$ & $68.87_{-0.53}^{+0.44}$ \\[0.1 cm]
		$100\,\omega_{\rm b}$        & $2.214_{-0.019}^{+0.031}$ & $2.229_{-0.016}^{+0.027}$ \\[0.1 cm]
		$\omega_{\rm cdm}$           & $0.12325_{-0.0054}^{+0.00092}$ & $0.11992_{-0.0026}^{+0.00066}$ \\[0.1 cm]
		$\xi$                          & $<0.15$ & $<0.067$ \\[0.1 cm]
		$\epsilon$                   & $<4.1\times10^{-3}$ & $<3.8\times10^{-3}$ \\[0.1 cm]
		$\sigma_8$                   & $0.8053_{-0.0087}^{+0.032}$ & $0.8100_{-0.0063}^{+0.017}$ \\[0.1 cm]
		$S_8=\sigma_8(\Omega_m/0.3)^{0.5}$ & $0.822_{-0.014}^{+0.021}$ & $0.8096_{-0.0094}^{+0.012}$ \\[0.1 cm]
		\hline
		\rule{0pt}{3.0ex}
		$\Delta\chi^2$               & $-0.8$ & $0.0$ \\[0.1 cm]
		$\sigma$             		 & $0.4$ & $0.0$ \\[0.1 cm]
		$\ln B_{\rm iDE\gamma,\Lambda}$          & $-9.7$ & $-11.6$ \\[0.1 cm]
		\hline
	\end{tabular}
	\caption{Mean and 68\% C.L. (the lower/upper bounds are given at the 95\% C.L.) of the parameters most significantly affected by the DM-DE-photon interactions for different data set combinations, together with the corresponding $\Delta\chi^2$, $\sigma$, and $\ln B_{\rm iDE\gamma,\Lambda}$ values.}	
	\label{tab: MCMC_full}
\end{table}

The quantitative evaluation of this possibility is shown in the first column of Tab. \ref{tab: MCMC_full}, where we consider the full iDMDE$\gamma$ model with $\xi$ assumed to be positive (we do not show the $\xi$-negative case for sake of brevity since it does not present any interesting cosmological feature, as we check explicitly). There, it is possible to observe that, although the upper limit on $\epsilon$ does significantly relax (by more than two orders of magnitude), the model's ability to solve the Hubble tension does not improve (but instead slightly worsens). The same is true also with respect to the $S_8$ tension when comparing the value reported in Tab.~\ref{tab: MCMC_full} to that listed in Tab. I of~\cite{Lucca:2021dxo}, i.e., ${S_8=0.813_{-0.014}^{+0.019}}$.

This behavior, in a qualitative way, appears to be most prominently due to the fact that the two interaction channels and their impact on the observables are not additive, but rather mildly competing (as discussed in Sec.~\ref{sec: obs}). In this way, although the iDE$\gamma$ and the iDMDE model can increase the $H_0$ value and decrease the $S_8$ value, respectively, at the same time, they prefer the $\Lambda$CDM value for the other parameter ($S_8$ and $H_0$, respectively), which ultimately spoils either model's ability to shift the $\Lambda$CDM prediction in a considerable manner. 

\begin{figure}[t]
	\centering
	\includegraphics[width=0.95\columnwidth]{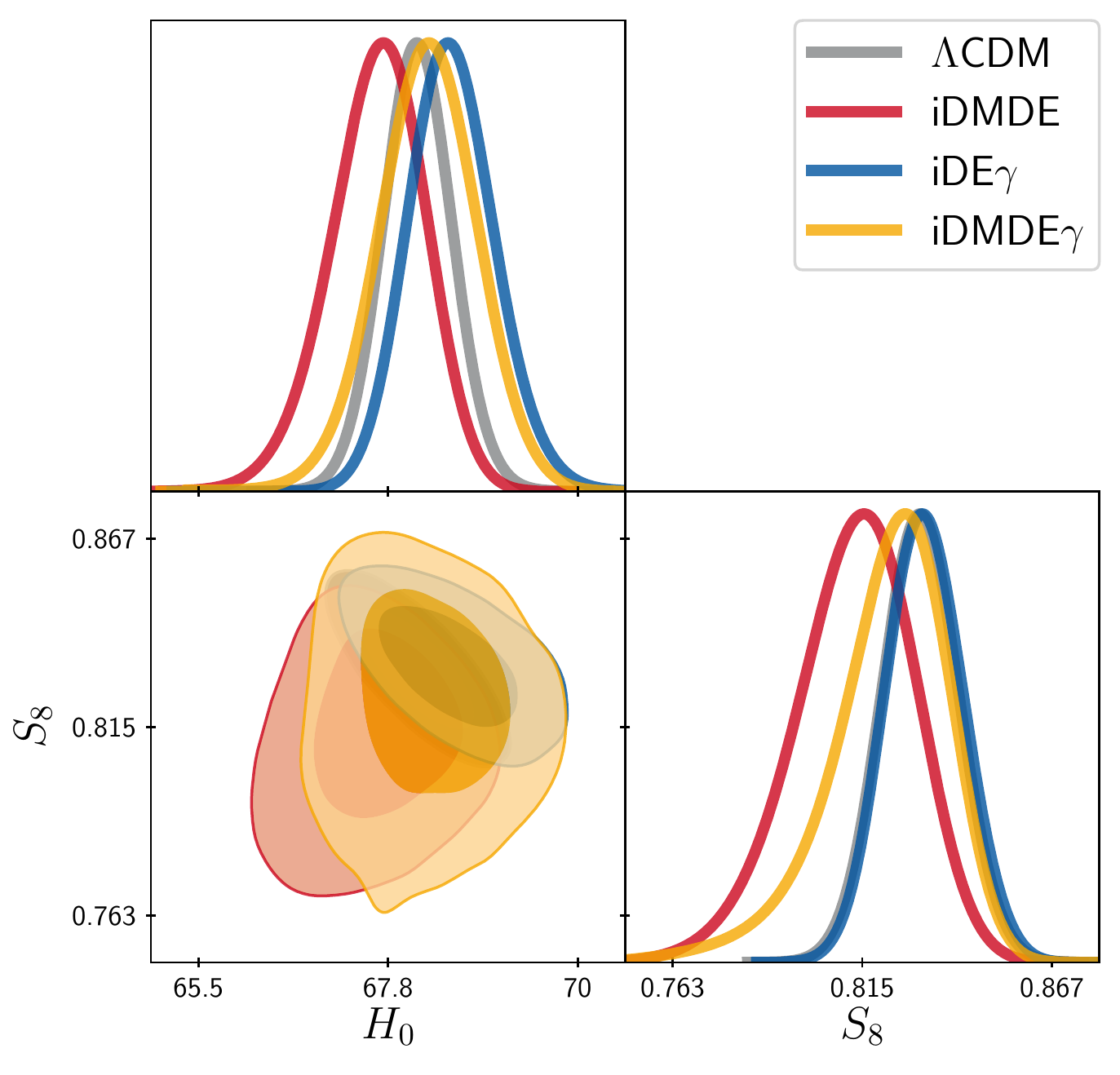}
	\caption{Two-dimensional contours (68\% and 95\% C.L.) and one-dimensional distributions of the $S_8-H_0$ plane for different models and assuming the \textit{baseline} data set combination.}
	\label{fig: MCMC_res}
\end{figure}

This limit of the model becomes even more transparent with the help of the posterior distributions shown in Fig.~\ref{fig: MCMC_res}. Indeed, there one can see that the orange contour, which represent the fully interacting iDMDE$\gamma$ model, does not correspond to the sum of the red (taken from \cite{Lucca:2021dxo})  and blue distributions, corresponding to the iDMDE and iDE$\gamma$ models, respectively, but always lays between them. As a consequence, on the one hand the allowed model's parameter space shrinks (as does its ability to restore the concordance between early- and late-time measurements), while on the other hand the mean values of the cosmological parameters involved always reduce to a weighted average of the ones preferred by the two interacting models considered singularly, inevitably reinforcing the agreement with $\Lambda$CDM. 

Nevertheless, despite this limitation, the fully interacting model can still reduce the significance of the $S_8$ tension between the inference of this parameter from the \textit{baseline} data sets ($S_8=0.8216_{-0.041}^{+0.034}$ at $2\sigma$) and its determination via weak lensing surveys such as KV450 ($S_8=0.737_{-0.036}^{+0.040}$~\cite{Hildebrandt:2018yau}) and DES ($S_8=0.782\pm0.027$~\cite{Troxel:2017xyo}), as also clear from Fig. \ref{fig: MCMC_res}. Therefore, it is in principle possible to combine these data sets (\textit{baseline} and KV450+DES, as explained in Sec.~\ref{sec: meth}) and we explore this possibility for sake of completeness. 

We show the resulting constraints from the \textit{baseline}+KV450+DES combination on the cosmological parameters most impacted by the iDMDE$\gamma$ model in the right column of Tab. \ref{tab: MCMC_full}. As expected, the $S_8$ value further decreases, to the point where the tension with respect to both KV450 and DES drops below the 2$\sigma$ level (1.9$\sigma$ and 1.0$\sigma$, respectively). Also the $H_0$ value shifts to slightly higher values, but the tension with the late-time estimate reported by the SH0ES collaboration \cite{Riess2019Large}, $H_0=74.03\pm1.42$ km/(s Mpc), only reduces to 3.4$\sigma$.

As a final, remark, for all models we considered we observe a strong Bayesian evidence in favor of $\Lambda$CDM (see Sec. \ref{sec: meth}). This is in large part due to the presence of the DE-photon interactions which are themselves significantly disfavored (see Tab. \ref{tab: MCMC_eps}), while the DE-DM interactions show only a negligible preference for $\Lambda$CDM (see Tab. I of \cite{Lucca:2021dxo}).

\section{Conclusions}\label{sec: conc}
It has been convincingly argued that \tquote{the least unlikely} cosmological model able to successfully address the Hubble tension has to modify the expansion history of the universe just prior to recombination. Nevertheless, significant issues have been identified with models that only do so, where secondary tensions can be introduced or exacerbated between early- and late-time observations (for instance involving the matter energy density or the $\sigma_8$ parameter, respectively). Therefore, there seems to be emerging the need for a model able to affect the standard $\Lambda$CDM model at multiple epochs: before recombination in order to restore the concordance between early-time inference and late-time measurements of the $H_0$ value, and at late times to compensate for eventual shifts in the $\Lambda$CDM parameters and/or to alleviate the $\sigma_8$ tension.

Following this philosophy, in this paper we consider a multi-interacting cosmological model where a fraction of the DE energy density can flow to the DM and to the photons via a coupling function proportional to the DE energy density. The reason behind the choice of these interaction channels can be understood following the time dependence of the two fluids the DE can interact with. Indeed, DE-DM interactions affect almost exclusively late times and have been shown to be able to successfully solve the $\sigma_8$ tension for the aforementioned choice of coupling function. On the other hand, DE-photon interactions are particularly relevant during radiation domination (i.e., only prior to recombination) and could in principle be able to alleviate the $H_0$ tension by introducing an additional and natural freedom in the expansion history of the universe just prior to recombination as well as by being able to accelerate photon decoupling (and ultimately impact the sound horizon).

After a very detailed description of how this model affects the expansion and thermal history of the universe and how this reflects on key observables such as the CMB temperature anisotropy power spectrum and the matter power spectrum, we tested the model's ability to successfully address the $H_0$ and $S_8$ tensions (i.e., without introducing any other tension nor exacerbating either one of them) considering data from \textit{Planck}, BAO and Pantheon. We find that the DE-photon interactions can indeed increase the $H_0$ value, but not to a degree where the concordance between early- and late-time observations is restored at a statistically significant level. The reason for this is most probably due to the fact that the modifications introduced by this interaction to the background and thermal history are too diluted in time, so that the following changes in the observables cannot be accommodated by the data nor reabsorbed by degeneracies with other parameters. This is compatible with the idea that for a solution to the $H_0$ tension to be successful it needs to modify the history of the universe only a few decades in redshift prior to recombination. 

However, due to the DE-DM interactions, the fully interacting model is able to solve the $\sigma_8$ tension and it is therefore in principle possible to include for sake of generality data from weak lensing surveys, such as KV450 and DES, to the \textit{baseline} data sets. By doing so, as a final result we obtain ${H_0=68.87_{-0.53}^{+0.44}}$~\text{km/(s Mpc)} and $S_8=0.8096_{-0.0094}^{+0.012}$, so that the model reduces the $H_0$ and $S_8$ tensions below 3.5$\sigma$ and 2$\sigma$, respectively (the former with respect to the value reported by SH0ES and the latter with respect to KV450 and DES). The statistical evidence is however strongly in favor of $\Lambda$CDM (regardless of the data set combination).

Overall, we conclude that this type of multi-interacting DE model, despite its broad generality and very rich cosmological phenomenology, is not able to successfully solve the $H_0$ and $S_8$ tensions at the same time.

\section*{Acknowledgements}
The authors thank Thomas Hambye for useful discussions during the whole development this work. ML is supported by an F.R.S.-FNRS fellowship, by the \tquote{Probing dark  matter with neutrinos} ULB-ARC convention and by the IISN convention 4.4503.15. Computational resources have been provided by the Consortium des Équipements de Calcul Intensif (CÉCI), funded by the Fonds de la Recherche Scientifique de Belgique (F.R.S.-FNRS) under Grant No. 2.5020.11 and by the Walloon Region.

\appendix
\section{Derivation of the background equations}\label{sec: app bg}
In this appendix we derive the energy density evolution equations for the DM, photons and DE.

Starting from the case of the DE, Eq. \eqref{eq: rho_x} can easily be rewritten as
\begin{align}
	\frac{{\rm d}\rho_x}{\rho_x}=-3(1+w_{{\rm eff},x})\frac{{\rm d}a}{a}\,,
\end{align}
which can immediately be solved to obtain Eq. \eqref{eq: rho_x 2}. The situation is slightly more complicated for the DM and the photons. In particular, in the case of the photons we start by rewriting Eq. \eqref{eq: rho_g} as
\begin{align}
	\nonumber \frac{{\rm d}\rho_\gamma}{\rho_\gamma} & =-4\frac{{\rm d}a}{a}+\epsilon \xi \frac{\rho_x}{\rho_\gamma}\frac{{\rm d}a}{a} \\
	& \simeq -4\frac{{\rm d}a}{a}+\epsilon \xi \frac{\rho_{x,0}}{\rho_{\gamma,0}}a^{-3w_{{\rm eff},x}+1}\frac{{\rm d}a}{a}\,,
\end{align}
where in the second equality we have made use of Eq.~\eqref{eq: rho_x 2} and assumed that at leading order ${\rho_\gamma\simeq\rho_{\gamma,0}a^{-4}}$. The solution of this differential equation (integrating between today, where $a=a_0=1$, and a given scale factor $a$) reads
\begin{align}\label{eq: rho_g/rho_g0}
	\nonumber \frac{\rho_\gamma}{\rho_{\gamma,0}} & =a^{-4}\exp\left[\epsilon \xi \frac{\rho_{x,0}}{\rho_{\gamma,0}}\int_{a_0}^{a} \frac{{\rm d}a}{a^{3w_{{\rm eff},x}-1+1}}\right] \\ \nonumber
	& =a^{-4}\exp\left[\frac{\epsilon \xi}{3w_{{\rm eff},x}-1} \frac{\rho_{x,0}}{\rho_{\gamma,0}}\right. \\ & \hspace{2.5 cm} \times \left.\left(1-a^{-3w_{{\rm eff},x}+1)}\right)\right]\,.
\end{align}
Assuming the argument of the exponential function to be small (which is the case given its proportionality to two constants which we know to be much smaller than 1, and since it scales approximately like $a^{4}$ with standard values), it is possible to Taylor expand this expression like $e^x\simeq 1+x$ to obtain Eq. \eqref{eq: rho_g 2}. Finally, in the same manner, we can solve Eq. \eqref{eq: rho_c} for the DM to find Eq. \eqref{eq: rho_c 2}.

Making use of the \texttt{ndf15} ODE solver implemented by default in CLASS \cite{Lesgourgues2011Cosmic, Blas2011Cosmic}, we numerically solve the coupled differential equations \eqref{eq: rho_c}-\eqref{eq: rho_x} to find that the analytic solutions derived above are indeed very accurate (as long as the product of $\xi$ and $\epsilon$ is small). 

\section{Derivation of the thermodynamics equations}\label{sec: app ther}

In this appendix we derive the evolution equation of the photons temperature starting from Eq. \eqref{eq: dot_t}, which was first derived in \cite{Lima:1995kd, Lima:2000ay}. For the first part, although with a slightly different approach, we will follow the derivation outlined in \cite{Jetzer:2011kw}, which we will then generalize letting $w_x$ be free and including also the contribution from the radiation.

We start by remarking that, in order to preserve a black body shape of the radiation energy spectrum over the whole thermal history of the universe, the second term in Eq. \eqref{eq: dot_t} has to vanish. This immediately implies that
\begin{align}\label{eq: from_BB}
	\frac{\psi_\gamma}{n_\gamma}=\frac{3\epsilon Q}{4\rho_\gamma}\,.
\end{align}
Then Eq. \eqref{eq: dot_t} simplifies to
\begin{align}
	\frac{\dot{T_\gamma}}{T_\gamma}= \left(\frac{\partial p_\gamma}{\partial \rho_\gamma}\right)_n\frac{\dot{n}_\gamma}{n_\gamma}=\frac{1}{3}\frac{\dot{n}_\gamma}{n_\gamma}
\end{align}
since $(\partial p_\gamma/\partial \rho_\gamma)_n=w_\gamma$ by definition. Using Eqs. \eqref{eq: dot_n}, \eqref{eq: from_BB} and \eqref{eq: C_x 2} one finds that
\begin{align}\label{eq: dot_T}
	\nonumber \frac{\dot{T_\gamma}}{T_\gamma} & = \frac{1}{3}\left(\frac{\psi_\gamma}{n_\gamma}-3H\right)= \frac{1}{3}\left(\frac{3\epsilon Q}{4 \rho_\gamma}-3H\right) \\ & = -H\left(1+\frac{\epsilon \xi}{4}\frac{\rho_x}{\rho_\gamma}\right)\,.
\end{align}
This leads to a differential equation of the form
\begin{align}\label{eq: dT/T}
	\frac{\text{d} T_\gamma}{T_\gamma} = -\left(1+\frac{\epsilon \xi}{4}\frac{\rho_x}{\rho_\gamma}\right)\frac{\text{d}a}{a}\,,
\end{align}
with solution
\begin{align}\label{eq: dT/T 2}
	\frac{T_\gamma}{T_{\gamma,0}} = a^{-1}\exp\left(-\frac{\epsilon \xi}{4}\int_{a_0}^{a}\frac{\rho_x}{\rho_\gamma}\frac{\text{d}a}{a}\right)\,.
\end{align}

Note that, although in a different form, this last equation is perfectly equivalent to Eq. (2.30) of \cite{Jetzer:2011kw} (assuming $\gamma=4/3$). However, in order to solve the integral in the exponential function, \cite{Jetzer:2011kw} made use of a simplified form of the Hubble rate evolution which neglected the role of radiation and assumed that all the matter was in form of DM. Although this approach is quite accurate for the redshifts considered in the analysis of the reference, here we go beyond these assumptions and solve this expression exactly making use of Eqs.~\eqref{eq: rho_g 2} and~\eqref{eq: rho_x 2}. Indeed, inserting these equations in Eq. \eqref{eq: dT/T 2} and with the identity
\begin{align}
	\int \frac{a^{-x-1}}{1+y(1+a^{-x})} \text{d}a=\frac{\ln(|y(a^{-x}-1)-1|)}{xy}\,,
\end{align}
where in our case we have that $x=3w_{{\rm eff},x}-1$ and ${y=\epsilon \xi \rho_{x,0}/((3w_{{\rm eff},x}-1)\rho_{\gamma,0})}$, one finds that
\begin{align}
	\nonumber \frac{T_\gamma}{T_{\gamma,0}} = & a^{-1}\left[1+\frac{\epsilon \xi}{3w_{{\rm eff},x}-1}\frac{\rho_{x,0}}{\rho_{\gamma,0}}\right. \\ & \hspace{1.4 cm} \times \left. (1-a^{-3w_{{\rm eff},x}+1}) \right]^{1/4}\,,
\end{align}
or equivalently Eq. \eqref{eq: T_g}.

As a remark, it is interesting to underlay the fact that Eq. \eqref{eq: dT/T 2} can be solved in the same way as Eq. \eqref{eq: rho_g/rho_g0}, i.e., with the same assumptions on $\rho_\gamma$ and Taylor expanding the exponential function, since they only differ in the numerical prefactor\footnote{This would deliver the same result as expanding Eq. \eqref{eq: dT/T 2} in the exponent and neglecting higher order terms.}. Indeed, also in this case, we confirm that the difference between these two solutions to Eq.~\eqref{eq: dT/T 2}, the exact and the approximated one, is negligible (as long as $\xi\epsilon$ is very small), so that we will use the exact solution for sake of generality but without breaking the consistency with the previous results.

\section{Derivation of the perturbation equations}\label{sec: app per}
In this appendix we derive the perturbation equation of the DM, photon and DE fluids coupled as in Eqs.~\text{\eqref{eq: rho_c}-\eqref{eq: rho_x}}. We will do so in the synchronous gauge and in a flat FLRW universe. When referring to the standard scenario, we will implicitly assume the equations outlined in e.g.,~\cite{Ma1995Cosmological}. We will also employ the same conventions as the reference, where the interested reader can find more details about the underlying perturbation theory (see also e.g., \cite{Lesgourgues:2013qba} for a complete pedagogical introduction). To describe the interacting fluids, it is in this context useful to rewrite the coupling function $Q$ in its four-component form \cite{Gavela2009Dark, Gavela2010Dark}, i.e., 
\begin{align}
	Q^\nu=\xi H\rho_x u_c^\nu\,,
\end{align}
where $u_c^\nu$ is the DM four-velocity. Expressed in this way, it shows that, as commonly done in the literature (see e.g., \cite{Gavela2009Dark, Gavela2010Dark} as well as \cite{Valiviita2008Large}, although the latter considers a proportionality on a different energy density), one can assume $Q^\nu$ to be parallel to $u_c^\nu$ which avoids any momentum transfer in the DM rest frame and circumvents fifth-force constraints. We will make this assumption here.

In full generality, a fluid's perturbation equations in the synchronous gauge read \cite{Ma1995Cosmological}
\begin{align}
	& \dot{\delta} =-(1+w)\left(\theta+ \frac{\dot{h}}{2}\right) -3H\left(\frac{\delta p}{\delta \rho}-w\right)\delta\,, \label{eq: dot_delta} \\
	& \dot{\theta}=-H\theta (1-3w) -\frac{\dot{w}}{1+w}+\frac{\delta p}{\delta \rho}\frac{k^2\delta}{1+w}-k^2\sigma\,, \label{eq: dot_theta}
\end{align}
where $\delta=\delta \rho/\rho$ with $\delta \rho$ as the first-order perturbation of the energy density $\rho$, $\theta$ is the fluid's divergence velocity, $h$ is the trace part of the metric perturbation, the ratio $\delta p/\delta \rho=c_s^2$ is the general definition of the speed of sound ($\neq w$ for imperfect fluids such as scalar fields), $k$ represents the Fourier modes and $\sigma$ is defined in Eq. (22) of~\cite{Ma1995Cosmological}.

In the case of the DM, which we choose as the reference frame as usual in the synchronous gauge, these equations simplify to
\begin{align}\label{eq: DM pert mid}
	\dot{\delta}_c =-\frac{\dot{h}}{2} \quad \text{and} \quad \dot{\theta}_c=0\,,
\end{align}
assuming that $w_c=\dot{w}_c=c_{s,c}^2=\delta p_c/\delta \rho_c=0$, as in the standard scenario, and employing $\theta_c=\sigma=0$ as the (synchronous) gauge choice. However, in addition to the terms already present in Eq. \eqref{eq: DM pert mid}, the perturbed energy density $\delta \dot{\rho}_c$ also gets a non-standard contribution $\delta Q_c$ because of the extra term in Eq. \eqref{eq: rho_c}, where ${Q_c=(1-\epsilon)Q}$. In terms of $\delta_c=\delta \rho_c/\rho_c$, this means that
\begin{align}\label{eq: delta_c mid}
	\dot{\delta}_c=[\dots]+\frac{\delta Q_c}{\rho_c}-\frac{\delta_c Q_c}{\rho_c}
\end{align}
where $[\dots]$ (now and henceforth) includes all of the standard terms which are already accounted for in Eqs. \eqref{eq: dot_delta}-\eqref{eq: dot_theta} and therefore do not need to be included again. In order to find $\delta Q_c$, it is necessary to rewrite the coupling term $Q$ of Eq. \eqref{eq: C_x 2} in terms of the diagonal terms of the metric (the only non-zero ones in a perturbed FLRW universe), so that we obtain
\begin{align}
	\delta Q =\xi\delta\left(\frac{\dot{g}_{ii}}{g_{ii}}\right)\rho_x+\xi\frac{\dot{g}_{ii}}{g_{ii}}\delta \rho_x = \xi\rho_x\left(\frac{\dot{h}}{6}+H\delta_x\right)\,.
\end{align}
Then, Eq. \eqref{eq: delta_c mid} can be solved to obtain
\begin{align}
	\dot{\delta}_c =[\dots]+(1-\epsilon)\xi\frac{\rho_x}{\rho_c}\left[\frac{\dot{h}}{6}+H(\delta_x-\delta_c)\right]\,.
\end{align}
Since we assume here to have zero momentum transfer between the fluids, no additional term has to be taken into account for $\dot{\theta}_c$.

Next, we focus our attention to the case of the photons and the DE. Here it is first of all important to remark the fact that, differently from for instance the adiabatic sound speed $c_a^2$ (which is solely determined by the EOS parameter of the fluid), the general speed of sound ${c_s^2=\delta p/\delta \rho}$ which enters in Eqs.~\eqref{eq: dot_delta}-\eqref{eq: dot_theta} is neither scale nor gauge independent (see e.g., Sec. II of \cite{Bean:2003fb} for a more in-depth discussion about this aspect). Therefore, it is fundamental to map the sound speed $c_s^2$ of a given fluid in the chosen rest frame, in our case the DM. This can be achieved with the transformation \cite{Bean:2003fb}
\begin{align}\label{eq: cs trafo}
	\hat{c}_s^2 = c_s^2+3H(1+w_{\rm eff})(c_s^2-c_a^2)\frac{\theta}{k^2\delta}\,,
\end{align}
where the $\,\hat{}\,$ characterizes quantities in a given fluid's frame and $w_{\rm eff}$ the effective EOS parameter of that fluid. In the particular case of photons we have
\begin{align}
	w_{\rm eff, \gamma}=w_\gamma - \frac{\epsilon Q}{3 H\rho_\gamma}=w_\gamma - \frac{\epsilon \xi}{3}\frac{\rho_x}{\rho_\gamma}\,,
\end{align}
while for the DE fluid it reads
\begin{align}
	w_{{\rm eff},x}=w_x + \frac{Q}{3 H\rho_x}=w_x + \frac{\xi}{3}\,.
\end{align}

Focusing firstly on the photons, one can then derive the general definition of the sound speed (see e.g., Eq.~(7.85) of~\cite{Mukhanov:2005sc} and related text)
\begin{align}\label{eq: cs}
	c_{s,\gamma}^2=\frac{1}{3}\left(1+\frac{3}{4}\frac{\rho_b}{\rho_\gamma}\right)^{-1}
\end{align}
to find that, if one neglects the contribution of the baryons as done in the standard case (and very accurate for the longest part of RD), $c_{s,\gamma}^2\simeq1/3$. As a consequence, we have that $c_{s,\gamma}^2\simeq c_{a,\gamma}^2= w_\gamma$, so that the second term in Eq.~\eqref{eq: cs trafo} vanishes and we do not need to account for changes in the photon sound speed due to our choice of reference frame (as in the standard case). However, similarly as for the DM, also in the case of photons we get an additional contribution to $\delta \dot{\rho}_\gamma$ from the term $\delta Q_\gamma$, where $Q_\gamma = \epsilon Q$. Then we have that
\begin{align}\label{eq: delta_g mid}
	\nonumber \dot{\delta}_\gamma & =[\dots]+\frac{\delta Q_\gamma}{\rho_\gamma}-\frac{\delta_\gamma Q_\gamma}{\rho_\gamma}\\
	& = [\dots] + \epsilon \xi \frac{\rho_x}{\rho_\gamma}\left[\frac{\dot{h}}{6}+H(\delta_x-\delta_\gamma)\right]\,,
\end{align}
where the $[\dots]$ refers to the standard terms already accounted for in Eq. (58) of \cite{Ma1995Cosmological}. Since the assumption of no momentum transfer is valid also for the photons, no additional contribution to $\dot{\theta}_\gamma$ needs to be accounted for due to the coupling to the DE. The same is true also for the momentum average phase-space density perturbation defined in Eq. (58) of \cite{Ma1995Cosmological}.

On the other hand, assuming for the DE that ${c_{s,x}^2=c_{a,x}^2=w_x}$ would inevitably lead to instabilities since the EOS parameter is negative and the sound speed $c_{s,x}$ would then be imaginary. It is therefore necessary to impose the condition $c_{s,x}^2>0$ and a common choice is to adopt the scalar field value $c_{s,x}^2=1$ (see Sec. 2.3 of \cite{Valiviita2008Large} for a derivation and additional details). In this way, the two sound speeds do not coincide and we have that
\begin{align}\label{eq: cs trafo DE}
	\nonumber\hat{c}_{s,x}^2 = & c_{s,x}^2+3H(1+w_x)  \left[1+\frac{\xi}{3 (1+w_x)}\right] \\ & \hspace{2.55 cm}\times (c_{s,x}^2-w_x)\frac{\theta_x}{k^2\delta_x}	
\end{align}
assuming $c_a^2=w_x$, as in the usual case. With this transformation the Eqs.~\text{\eqref{eq: dot_delta}-\eqref{eq: dot_theta}} for the DE become
\begin{align}
	& \nonumber \dot{\delta}_x =-(1+w_x)\left(\theta_x+ \frac{\dot{h}}{2}\right) \\ &\hspace{.7 cm} -3H\left(c_{s,x}^2-w_x\right)\left[\delta_x+\frac{H\theta_x}{k^2}(3(1+w_x)+\xi)\right]\,, \label{eq: dot_delta_x} \\
	& \nonumber \dot{\theta}_x=-H\theta_x (1-3w_x) +\frac{c_{s,x}^2 k^2\delta_x}{1+w_x} \\ & \hspace{0.7 cm} +3H\theta_x(c_{s,x}^2-w_x)\left[1+\frac{\xi}{3(1+w_x)}\right]\,. \label{eq: dot_theta_x}
\end{align}
As a final step, we need to ensure energy and momentum conservation. The former is obtained by imposing the condition
\begin{align}
	\nonumber \delta\dot{\rho}_x+ (\delta\dot{\rho}_c+ \delta\dot{\rho}_\gamma) & =0 \\
	& \nonumber =\delta\dot{\rho}_x +[\dots]+ \delta Q \\
	& =\dot{\rho}_x \delta_x + \dot{\delta}_x\rho_x +[\dots]+ \delta Q\,,
\end{align}
where in the second equality we made use of the fact that the only non-standard contributions to $\delta\dot{\rho}_c$ and $\delta\dot{\rho}_\gamma$ are $\delta Q_c$ and $\delta Q_\gamma$, respectively, which sum up to $\delta Q$. This can then be solved in terms of $\dot{\delta}_x$ to have
\begin{align}
	\dot{\delta}_x=[\dots]+\xi H\delta_x-\xi\left(\frac{\dot{h}}{6}+H\delta_x\right)=[\dots]-\xi\frac{\dot{h}}{6}\,.
\end{align}
Finally, momentum conservation is found making use of the condition
\begin{align}
	\frac{d}{dt}[(p_c+\rho_c)\theta_c+(p_\gamma+\rho_\gamma)\theta_\gamma+(p_x+\rho_x)\theta_x]=0\,,
\end{align}
from which follows that
\begin{align}
	\nonumber \dot{\theta}_x & =-\frac{\dot{\rho}_x\theta_x}{\rho_x}-\frac{\dot{\rho}_c\theta_c}{(1+w_x)\rho_x}-\frac{4}{3}\frac{\dot{\rho}_\gamma\theta_\gamma+\rho_\gamma\dot{\theta}_\gamma}{(1+w_x)\rho_x} \\
	& = [\dots]+\frac{\xi H}{1+w_x}[(1+w_x)\theta_x-(1-\epsilon)\theta_c -\gamma\epsilon \theta_\gamma]\,,
\end{align}
where we have neglected the term proportional to $\dot{\theta}_\gamma$ since, as mentioned before, in this case no non-standard contribution has to be taken into account because of the assumption of no momentum exchange between the fluids. Here the $[\dots]$ refer to Eq. \eqref{eq: dot_theta_x} and together the two equations can be rewritten as in Eq. \eqref{eq: dot theta_x 2}.

Concerning the initial conditions, they are unmodified for the DM and the photons, while those of the DE fluid can be shown to be \cite{Salvatelli2013New}
\begin{align}
	\delta_{x}^{i n}(x)=(1+w_x+2 \xi) C \quad \text { and } \quad \theta_{x}^{i n}=k^{2} \tau C\,,
\end{align}
with
\begin{align}
	C=-\frac{1+w-\xi/3}{12 w_x^{2}-2 w_x+3 w_x \xi-7 \xi-14} \frac{2 \delta_{\gamma}^{i n}}{1+w_{\gamma}}\,,
\end{align}
assuming adiabatic initial conditions for all other fluids, as it is the case in the scenario considered here.

\newpage
\bibliography{bibliography}{}

\end{document}